\documentclass[12pt]{revtex4}

\usepackage{multirow}
\usepackage{amsmath,amsfonts}
\usepackage{amssymb,latexsym}
\usepackage{color}
\usepackage{graphicx}
\usepackage{mathrsfs}

\newtheorem{prop}{Proposition}

\newcommand{\beprop}{\begin{prop}}
\newcommand{\enprop}{\end{prop}}
\newcommand{\bprf}{\begin{proof}} 
\newcommand{\eprf}{\end{proof}\qed}

\definecolor{hervecolor}{rgb}{0.8,0,0.7}

\newcommand{\greekbf}[1]{\mbox{\small{\boldmath$#1$}}}

\newcommand{\ket}[1]{|\kern.3ex#1\kern.3ex\rangle}
\newcommand{\bra}[1]{\langle\kern.3ex #1 \kern.3ex|}
\newcommand{\scalar}[2]{\langle\kern.3ex #1 \kern.3ex|\kern.3ex#2\kern.3ex\rangle}

\newcommand{\ii}{\mathsf{i}}

\newcommand{\ud}{\mathrm{d}}

\newcommand{\uE}{\mathrm{E}}

\begin{document}

\title{Perturbative expansion of irreversible works in symmetric and asymmetric processes}
\author{T.\ Koide}
\begin{abstract}
The systematic expansion method of the solution of the Fokker-Planck equation is developed by generalizing the formulation proposed in [J. Phys. A50, 325001 (2017)]. 
Using this method, we obtain a new formula to calculate the mean work perturbatively which is applicable to systems with degeneracy in the eigenvalues of the Fokker-Planck operator. 
This method enables us to study how the geometrical symmetry affects thermodynamic description of Brownian motion. 
To illustrate the application of the derived theory, we consider the Fokker-Planck equation with a two-dimensional harmonic potential. 
To investigate the effect of symmetry of the potential, we study thermodynamic properties in symmetric and asymmetric deformation processes of the potential: the rotational symmetry of the harmonic potential is held in the former, but it is broken in the latter. 
Optimized deformations in these processes are defined by minimizing mean works. 
Comparing these optimized processes, we find that the difference between the symmetric and asymmetric processes is maximized when the deformation time of the potential is given by a critical time which is characterized by the relaxation time of the Fokker-Planck equation. 
This critical time in the mean work is smaller than that of the change of the mean energy because of the hysteresis effect in the irreversible processes.
\end{abstract}

\maketitle

\section{Introduction}

There is no established framework to generalize the thermodynamic description to a small-fluctuating system, but 
such a system is often modeled using Brownian motion confined in an external potential 
\cite{hanngi-review,broeck-review,sekimoto-book,sei-rev}.
Then we can define mean work, heat and entropy in the process induced by 
the deformation of the confinement potential, 
and show that these quantities satisfy the laws analogous to the first and second laws in thermodynamics.
The deformation of the potential is implemented within a finite time period and thus any process is irreversible.
In the construction of, for example, an efficient nanomachines, 
it is important to find the optimized process where the irreversible contribution in mean works 
is minimized 
\cite{sekimoto-book,sekimoto-sasa,schmiedl-seifert2007,schmiedl-seifert2008,gomez,koning,aurell,geiger,then,dechant,bona,zulkow,crook,all,koide17}.

When the effect of inertia can be ignored, 
the distribution function of a Brownian particle in the configuration space is described by the Fokker-Planck equation.
Therefore the thermodynamic quantities in a small-fluctuating system can be obtained by solving it.
In order to find the optimized process, then, it is desirable to calculate 
the mean work in an analytic form \cite{sekimoto-sasa,sekimoto-book} 
because it is difficult to calculate variation numerically. 
In fact, the optimization has been exclusively studied for exactly solvable models like the harmonic potential  
\cite{schmiedl-seifert2007,schmiedl-seifert2008,gomez,speck,mazon,raybov,imparato,cohen,kwon,hol2,koide17,hol}. 
As applications to more general potentials, see, for example, \cite{blickle,dieterich}.
In Ref.\ \cite{koide17}, the present author developed a perturbative expansion method 
to calculate the solution of the Fokker-Planck equation in the one-dimensional system.
Using this theory, a formula to calculate the mean work perturbatively was obtained. 
Applying the derived perturbation theory to the one-dimensional harmonic potential, 
we confirmed that our perturbative calculations are consistent with the exact results.

In Ref.\ \cite{koide17}, however, the effect of degeneracy is not considered and thus the theory is 
not applicable to study the systems in higher spatial dimensions.
The purpose of this paper is to generalize the perturbation theory to arbitrary spatial dimensions.
Such a generalization enables us to study how the geometrical symmetry of the external potential affects thermodynamic description of a Brownian particle.
To illustrate the application of the derived theory, we study thermodynamic description of a Brownian particle confined 
in a two-dimensional harmonic potential. 
We are interested in the effect of symmetry and hence investigate symmetric and asymmetric compression processes. 
The rotational symmetry of the harmonic potential is held in the former, but it is broken in the latter.
Perturbative calculations are affected by these deformation processes because 
the degeneracy of the expansion basis (foot and head states) depends on the symmetry of the potentials. 
The optimized processes are found by minimizing the mean works.
Comparing the optimized symmetric and asymmetric processes, 
we find that these behaviors are the same in the vanishing and infinity limit of the deformation time of the potential.
The difference between these processes is maximized when the deformation time is given by a critical time 
which is characterized by the relaxation time of the Fokker-Planck equation. 
This critical time in the mean work is smaller than that in the change of the mean energy because of the hysteresis effect in irreversible processes.

This paper is organized as follows.
In Sec.\ \ref{sec:se}, the thermodynamic interpretation in the Fokker-Planck equation is summarized following stochastic energetics \cite{sekimoto-book}. 
In Sec.\ \ref{sec:eigenfunctions}, the eigenvalues and eigenfuctions of the 
time-dependent Fokker-Planck operator are introduced. 
Using these, we develop the perturbation theory in the D-dimensional Fokker-Planck equation 
in Sec.\ \ref{sec:pertur-fp}.
This result is applied to find the perturbative formula to calculate the mean work in Sec.\ \ref{sec:formula_mw}. 
To illustrate the applications of the formula, we consider symmetric and asymmetric compression processes of the two dimensional harmonic potential 
in Secs.\ \ref{sec:sym_prot} and \ref{sec:asym_prot}, respectively.
These results are numerically calculated and compared in Sec.\ \ref{sec:numerical}.
Section \ref{sec:concl} is devoted to the concluding remarks.

\section{Thermodynamic interpretation in Fokker-Planck equation} \label{sec:se}

Before discussing the perturbation theory, we briefly summarize the thermodynamic description 
of the systems described by the Fokker-Planck equation. 
The discussion in this section is based on stochastic energetics \cite{sekimoto-book}.

We consider a $D$-dimensional Brownian particle which is confined in an external potential $V$ and interacts with a thermal bath with temperature $T$. 
When the effect of friction is sufficiently large and hence that of inertia is negligibly small,  
the distribution of a Brownian particle in the configuration space, $\rho({\bf x},t)$, is described by  
the Fokker-Planck equation,
\begin{eqnarray}
\partial_t \rho ({\bf x},t) 
= \nabla \cdot \left\{ \frac{1}{\nu \beta} \nabla  
+ \frac{1}{\nu}  (\nabla V({\bf x} ,{\bf a}_t) ) \right\}  \rho ({\bf x},t) 
\equiv {\cal L}_t ({\bf x}) \rho({\bf x},t)\, . \label{eqn:fp}
\end{eqnarray}
where $\nu$ is the coefficient of friction and $\beta = 1/(k_B T)$ with $k_B$ being the Boltzmann constant.
In this work, the temperature is constant but the form of the potential $V$ is deformed 
by changing the control parameters ${\bf a}_t = (a^{(1)}_t,a^{(2)}_t, \cdots)$.
Because of the confinement potential $V$, the particle distribution $\rho ({\bf x},t)$ vanishes quickly 
at an infinite distance, ${\displaystyle \lim_{|{\bf x}| \rightarrow \infty}} \rho({\bf x},t) = 0$.

As is well-known, the Fokker-Planck equation (\ref{eqn:fp}) is reproduced from Brownian motion which is described by the stochastic differential equation,
\begin{eqnarray}
\ud {\bf x}_t  = -\frac{\ud t}{\nu} \nabla V ({\bf x}_t, {\bf a}_t) + \sqrt{\frac{2}{\beta \nu}} \ud {\bf W}_t \, .
\label{eqn:sde}
\end{eqnarray}
where 
$\ud {\bf W}_t= {\bf W}(t+dt)-{\bf W}(t)$ is the inclination of the standard Wiener process ${\bf W}(t)$ satisfying 
\begin{eqnarray}
\uE [\ud {W}^i_t] &=& 0 \, ,\\
\uE [\ud {W}^i_t \ud {W}^j_{t^\prime}] &=& \ud t \, \delta_{t,t^\prime} \, \delta_{i,j} \, .
\end{eqnarray}
The ensemble average of the Wiener process is denoted by $\uE [\,\,\,\,]$. 
See, for example, Ref.\ \cite{gardiner} for detail.
Using this stochastic trajectory of the Brownian particle, 
the particle distribution is defined by
\begin{eqnarray}
\rho({\bf x},t) = \int \ud {\bf R}\, \rho_0 ({\bf R}) \uE \left[ \delta^{(D)} ({\bf x}-{\bf x}_t) \right] \, ,
\end{eqnarray}
where $D$ is the spatial dimension and $\rho_0 ({\bf R})$ denotes the initial distribution of the Brownian particle at the initial position ${\bf R}$.
It is straightforward to show that $\rho({\bf x},t)$ satisfies the Fokker-Planck equation (\ref{eqn:fp}) \cite{gardiner,sekimoto-book}.

In this model, the Brownian particle interacts with the thermal bath through 
the dissipative force $(-\nu \ud {\bf x}_t/\ud t)$ and the random force $(\sqrt{{2}/{\beta \nu}}  \, {\ud {\bf W}_t}/{\ud t})$.
In stochastic energetics, the heat $\ud q_t$ associated with a single stochastic trajectory 
is interpreted as the work done by these interactions and thus given by 
\begin{eqnarray}
\ud q_t = \sum_{i=1}^D
\left(
- \nu \frac{\ud {x}^i_t}{\ud t } + \sqrt{\frac{2}{\beta \nu}}  \frac{\ud {W}^i_t}{\ud t}
\right) \circ \ud {x}^i_t 
 \, , \label{def:heat}
\end{eqnarray}
where $\circ$ is the Stratonovich definition of the product,
\begin{eqnarray}
{A}_t \circ \ud {W}_t = \frac{{A}_{t} + {A}_{t+\ud t}}{2}\ud {W}_t \, .
\end{eqnarray}
Using Eq.\ (\ref{eqn:sde}) to the definition (\ref{def:heat}), we can show that the above heat is reexpressed as
\begin{eqnarray}
\ud {q}_t 
=
\ud V ({\bf x}_t,{\bf a}_t) - \ud {w}_t \, . \label{eqn1stlaw_se}
\end{eqnarray}
The first term on the right-hand side gives the change of the internal energy 
and the second term represents the work done by an external system, which is induced by the deformation of the external potential and defined by
\begin{eqnarray}
\ud {w}_t = \sum_{i=1}^D \frac{\partial V ({\bf x}_t,{\bf a}_t)}{\partial a^i_t} \ud a^i_t \, .
\end{eqnarray}
Note that the relation (\ref{eqn1stlaw_se}) is satisfied for each stochastic trajectory

The expectation value of Eq.\ (\ref{eqn1stlaw_se}) leads to a law analogous to the first law of thermodynamics,
\begin{eqnarray}
\ud Q_t = \{ E({t+\ud t}) - E(t) \} - \ud W_t \, , \label{eqn:se_1law}
\end{eqnarray}
where
\begin{eqnarray}
\ud Q_t &=& \int \ud^D {\bf R} \, \rho_0 ({\bf R}) \uE[ \ud q_t ] \, ,\\
E(t) &=& 
\int \ud^D {\bf R} \, \rho_0 ({\bf R}) \uE[ V({\bf x}_t, {\bf a}_t) ]
=\int \ud^D {\bf x}\, \rho({\bf x},t) V({\bf x},{\bf a}_{t}) 
\label{eqn:def_ener} \, , \\
\ud W_t &=& 
\int \ud^D {\bf R} \, \rho_0 ({\bf R}) \uE[ \ud w_t ]
= \ud t \int \ud^D {\bf x} \, \rho({\bf x},t)   \sum_{i=1}^D \frac{\partial V ({\bf x},{\bf a}_t)}{\partial a^i_t} \frac{\ud a^i_t}{\ud t}  \, . \label{eqn:meanwork}
\end{eqnarray}

Using the Shannon entropy (multiplying the Boltzmann constant) defined by 
\begin{eqnarray}
S(t) = - k_B \int \ud^D {\bf x}\, \rho({\bf x},t) \ln \rho({\bf x},t) \, , \label{eqn:shannon}
\end{eqnarray}
we further find that there exists the following inequality \cite{sekimoto-book}: 
\begin{eqnarray}
\frac{\ud S(t)}{\ud t} - \frac{1}{T} \frac{\ud Q_t}{\ud t} \ge 0 \, , \label{eqn:2ndlaw}
\end{eqnarray}
where the equality is satisfied for the stationary state of the Fokker-Planck equation. 
This inequality can be regarded as the second law of thermodynamics.
Note, however, that the Shannon entropy is definable even in non-equilibrium states and thus 
the above inequality is not exactly the same as the the second law of thermodynamics.

In this paper, we consider the irreversible processes where the system interacts with the thermal bath with a fixed temperature $T$, 
and thus the mean work should satisfy the following inequality:
\begin{eqnarray}
\ud W_t = \{ E({t+\ud t}) - E(t) \}  - \ud Q_t \ge F_{helm} ({\bf a}_{t+\ud t}) - F_{helm} ({\bf a}_{t}) \, .
\label{eqn:vari_helm}
\end{eqnarray}
This is obtained by using the inequality (\ref{eqn:2ndlaw}) and 
the Helmholtz free energy defined by 
\begin{eqnarray}
F_{helm} ({\bf a}_{t}) = E(t) - T S(t) \, .\label{eqn:helmholtz}
\end{eqnarray}
Here, however, $S(t)$ is not the thermodynamic entropy but the Shannon entropy introduced above.
Therefore this Helmholtz free energy is not the same as the corresponding quantity in thermodynamics.

\section{Eigenvalues and eigenfunctions of time-dependent Fokker-Planck operator} \label{sec:eigenfunctions}

In this section, we define the expansion basis following Ref.\ \cite{koide17}. 
A similar expansion basis is discussed, for example, in Ref.\ \cite{gardiner}.  
We generalize this method to the case of the time-dependent Fokker–Planck operator referring to the discussion in Ref.\ \cite{namiki}.
Note that the eigenvalue theory of the time-periodic Fokker-Planck operator (the Kolmogorov operator) 
is discussed in Ref.\ \cite{caceres} and the properties found below are consistent with the result. 
See also Ref.\ \cite{risken}.

The eigenvalues and eigenfunctions of the time-dependent Fokker-Planck operator ${\cal L}_t ({\bf x})$ 
are defined by 
\begin{eqnarray}
{\cal L}_t ({\bf x})\,  \underline{\rho}_{n,\greekbf{\alpha}_n} ({\bf x},{\bf a}_t) = -  \underline{\lambda}_n (t)\, \underline{\rho}_{n,\greekbf{\alpha}_n} ({\bf x},{\bf a}_t) \, .
\end{eqnarray}
We call these eigenfunctions the {\it foot} states.
Here the index $n$ characterizes the eigenvalue $\underline{\lambda}_n (t)$ which is a function of time. 
The degeneracy associated with the index $n$  
is characterized by a set of non-negative integers $\greekbf{\alpha}_n = (\alpha_{1,n}, \alpha_{2,n}, \cdots)$, satisfying 
$0 \le \alpha_{i,n} \le d^{(i,n)}({\bf a}_t)$.
Thus the upper limit of the sums of degeneracy is characterized by 
a set of integers ${\bf d}^{(n)}({\bf a}_t) =(\ud^{(1,n)}({\bf a}_t), \ud^{(2,n)}({\bf a}_t),\cdots )$.
If there is no degeneracy, ${\bf d}^{(n)}({\bf a}_t) = (0, 0, \cdots) = {\bf 0}$, 
$\alpha_{i,n}$ can take only zeros, $\greekbf{\alpha}_n = (0, 0, \cdots) = {\bf 0}$.

Note that ${\cal L}_t ({\bf x})$ is not self-adjoint and the above eigenfunctions do not form a complete set. 
To find the complete set, we define the adjoint operator of ${\cal L}_t ({\bf x})$ by 
\begin{eqnarray}
\int \ud^D {\bf x} \, g ({\bf x})  {\cal L}_t ({\bf x}) f({\bf x}) = \int \ud^D {\bf x} \, \left(  {\cal L}^\dagger_t ({\bf x}) g ({\bf x}) \right)  f({\bf x}) \, ,
\end{eqnarray}
where $f({\bf x})$ and $g({\bf x})$ are smooth arbitrary functions and 
\begin{eqnarray}
{\cal L}^\dagger_t ({\bf x}) = 
\left[ \frac{1}{\nu \beta} \nabla^2  - \frac{1}{\nu}  (\nabla V({\bf x},{\bf a}_t) )\cdot \nabla \right] \, .
\end{eqnarray}
The eigenfunctions of  ${\cal L}^\dagger_t ({\bf x})$ are called the {\it head} states and defined by
\begin{eqnarray}
{\cal L}^\dagger_t ({\bf x}) \overline{\rho}_{n,\greekbf{\alpha}_n} ({\bf x},{\bf a}_t) = - \overline{\lambda}_n (t) \overline{\rho}_{n,\greekbf{\alpha}_n} ({\bf x},{\bf a}_t) \, .
\end{eqnarray} 
As shown soon later, the eigenvalue of the head state $\overline{\lambda}_n (t)$ 
is the same as that of the foot state $\underline{\lambda}_n (t)$.

Note that both of ${\cal L}_t({\bf x})$ and ${\cal L}^\dagger_t({\bf x})$ are characterized 
by a self-adjoint operator ${\cal H}_t({\bf x})$ defined by
\begin{eqnarray}
e^{G({\bf x},{\bf a}_t)/2} {\cal L}_t ({\bf x}) e^{-G({\bf x},{\bf a}_t)/2} 
= 
e^{-G({\bf x},{\bf a}_t)/2} {\cal L}^\dagger_t ({\bf x}) e^{G({\bf x},{\bf a}_t)/2} 
= - {\cal H}_t ({\bf x}) \, ,
\label{eqn:l-h}
\end{eqnarray}
where
\begin{eqnarray}
{\cal H}_t ({\bf x})
&=& \left[ -\frac{1}{\nu\beta} \nabla^2 - \frac{1}{2\nu} (\nabla^2 V({\bf x},{\bf a}_t) )
 + \frac{\beta}{4\nu} (\nabla V({\bf x},{\bf a}_t))^2 \right], \\
G({\bf x},{\bf a}_t) 
&=& \beta V({\bf x},{\bf a}_t) \, .
\end{eqnarray}
To give the representations of the foot and head states, we introduce the eigenfunctions of ${\cal H}_t ({\bf x})$,
\begin{eqnarray}
{\cal H}_t ({\bf x}) u_{n,\greekbf{\alpha}_n} ({\bf x},{\bf a}_t) = \lambda_{n}(t) u_{n,\greekbf{\alpha}_n} ({\bf x},{\bf a}_t) \, .
\end{eqnarray}
These eigenfunctions form a complete orthogonal set,
\begin{eqnarray}
\begin{split}
\int \ud^D {\bf x}\,  u_{n,\greekbf{\alpha}_n} ({\bf x},{\bf a}_t) u_{m,\greekbf{\beta}_m} ({\bf x},{\bf a}_t) 
= \delta_{n,m}\delta_{\greekbf{\alpha}_n, \, \greekbf{\beta}_n}, \\
\sum_{n\ge 0} \sum_{\greekbf{\alpha}_n={\bf 0}}^{{\bf d}^{(n)}({\bf a}_t)}  u_{n,\greekbf{\alpha}_n} ({\bf x},{\bf a}_t) u_{n,\greekbf{\alpha}_n} ({\bf x}',{\bf a}_t) = \delta^{(D)} ({\bf x} - {\bf x}')\, .
\end{split}
\label{eqn:compset_her}
\end{eqnarray}
where $\delta_{\greekbf{\alpha}_n, \greekbf{\beta}_n} 
= \delta_{\alpha_{1n},\beta_{1,n}} \, \delta_{\alpha_{2,n},\beta_{2,n}} \, \cdots$.
Note that the eigenvalues $\lambda_n (t)$ are non-negative real numbers as shown later 
and thus $u_{n,\greekbf{\alpha}_n}({\bf x},{\bf a}_t)$ can be chosen to be a real function.

Using these results, we construct the foot and head states.
The set of the foot and head states form a bi-orthogonal system and satisfies 
the following orthogonal and completeness conditions:  
\begin{eqnarray}
\begin{split}
\int \ud^D {\bf x}\ \overline{\rho}_{n,\greekbf{\alpha}_n} ({\bf x},{\bf a}_t) \underline{\rho}_{m,\greekbf{\beta}_m} ({\bf x},{\bf a}_t) 
= \delta_{n,m}\delta_{\greekbf{\alpha}_n,\greekbf{\beta}_m} \, , \\
\sum_{n\ge 0} \sum_{\greekbf{\alpha}_n ={\bf 0}}^{{\bf d}^{(n)}({\bf a}_t)}  
\underline{\rho}_{n,\greekbf{\alpha}_n} ({\bf x},{\bf a}_t) \overline{\rho}_{n,\greekbf{\alpha}_n} ({\bf x}',{\bf a}_t) = \delta^{(D)} ({\bf x} - {\bf x}') \, .
\end{split}
\label{eqn:nor_ori}
%\label{eqn:orth_ori}
\end{eqnarray}
Because of these properties, the solution of the Fokker-Planck equation can be expanded using the foot and head states.
Using the eigenfunctions of ${\cal H}_t$, the foot and head states are represented by  
\begin{eqnarray}
\begin{split}
\underline{\rho}_{n,\greekbf{\alpha}_n} ({\bf x},{\bf a}_t) 
&=& \sqrt{{\cal N}_{n,\greekbf{\alpha}_n}({\bf a}_t)  } 
 e^{-G({\bf x},{\bf a}_t)/2}u_{n,\greekbf{\alpha}_n} ({\bf x},{\bf a}_t) \, , \\
\overline{\rho}_{n,\greekbf{\alpha}_n} ({\bf x},{\bf a}_t) 
&=& \frac{1}{\sqrt{{\cal N}_{n,\greekbf{\alpha}_n}({\bf a}_t)}} 
e^{G({\bf x},{\bf a}_t)/2} u_{n,\greekbf{\alpha}_n} ({\bf x},{\bf a}_t)  \, . 
\end{split}
\label{eqn:urho_u}
%\label{eqn:orho_u}
\end{eqnarray}
Here we introduced 
the real factor ${\cal N}_{n,\greekbf{\alpha}_n}({\bf a}_t)$.
It is because, differently from quantum mechanics, the normalization condition is applied to a pair of foot and head states 
and thus, for example, a foot state itself is not necessarily normalized by one. 
This factor is not considered in the formulation of Ref.\ \cite{koide17}.

From Eqs.\ (\ref{eqn:urho_u}), we find that all eigenvalues of  
$\underline{\rho}_{n,\greekbf{\alpha}_n} ({\bf x},{\bf a}_t) $, $\overline{\rho}_{n,\greekbf{\alpha}_n} ({\bf x},{\bf a}_t) $ 
and $u_{n,\greekbf{\alpha}_n} ({\bf x},{\bf a}_t) $ are equal,
\begin{eqnarray}
 \lambda_n  (t) = \overline{\lambda}_n (t) = \underline{\lambda}_n (t)  \label{eqn:lam=ov=un} \, .
\end{eqnarray}
Moreover, the foot and head states satisfy the relation: 
\begin{eqnarray}
 \overline{\rho}_{n,\greekbf{\alpha}_n} ({\bf x},{\bf a}_t)
=
\frac{1}{{\cal N}_{n,\greekbf{\alpha}_n}({\bf a}_t)}e^{G({\bf x},{\bf a}_t)} \underline{\rho}_{n,\greekbf{\alpha}_n} ({\bf x},{\bf a}_t)  \, .
\label{eqn:ou1}
\end{eqnarray}
Therefore it is easy to understand that the degeneracy of $\underline{\rho}_{n,\greekbf{\alpha}_n} ({\bf x},{\bf a}_t)$ 
is the same as that of $\overline{\rho}_{n,\greekbf{\alpha}_n} ({\bf x},{\bf a}_t)$.

For an arbitrary external potential, 
we can show the following properties:
\begin{enumerate}

  \item The smallest eigenvalue is given by zero. Thus, without loss of generality, we can set the order of the eigenvalues as
\begin{equation}
0 = \lambda_0 (t)< \lambda_1 (t)< \lambda_2 (t)< \cdots \, .
\end{equation}

  \item The ``ground" foot and head states $\underline{\rho}_{0,{\bf 0}} ({\bf x},{\bf a}_t)$ and $\overline{\rho}_{0,{\bf 0}} ({\bf x}, {\bf a}_t)$ have the eigenvalue $\lambda_0(t)=0$. These states are not degenerate and given by 
\begin{eqnarray}
\begin{split}
\underline{\rho}_{0,{\bf 0}} ({\bf x},{\bf a}_t) &= \sqrt{{\cal N}_{0,{\bf 0}}({\bf a}_t)}
\frac{1}{\sqrt{{\cal Z} ({\bf a}_t)}} e^{-\beta V({\bf x},{\bf a}_t)} \, , \\
\overline{\rho}_{0,{\bf 0}} ({\bf x},{\bf a}_t) &= \frac{1}{\sqrt{{\cal N}_{0,{\bf 0}}({\bf a}_t)}}
\frac{1}{\sqrt{{\cal Z}({\bf a}_t)}} \, , 
\end{split}
\end{eqnarray}
respectively. 
Here we introduced 
\begin{equation}
{\cal Z} ({\bf a}_t) = \int \ud^D {\bf x} \, e^{-\beta V({\bf x},{\bf a}_t)} \, . \label{eqn:def_z}
\end{equation}

\end{enumerate}  
The derivations of these properties are the same as those in one dimension \cite{koide17} and 
are summarized in Appendix \ref{app:1}.

When we choose 
\begin{eqnarray}
{\cal N}_{0,{\bf 0}}({\bf a}_t) = \frac{1}{{\cal Z}({\bf a}_t)} \, ,
\end{eqnarray}
the ground head state becomes trivial
\begin{eqnarray}
\overline{\rho}_{0,{\bf 0}} ({\bf x},{\bf a}_t)  = 1 \, .
\end{eqnarray}
This choice is considered in Ref.\ \cite{koide18}, where 
it is shown that the time-evolution of the Fokker-Planck, the Kramers and relativistic Kramers systems can be regarded 
as a special case of Schr\"{o}dinger's reciprocal process 
and then the Jarzynski relation is reproduced from the symmetry of the Fokker-Planck operator.

\subsection{Bra-ket representation}

For the sake of simplicity, we introduce the quantum-mechanical bra-ket notation.
The foot and heat states are expressed in terms of the inner products of bra-ket vectors, 
\begin{eqnarray}
\begin{split}
\underline{\rho}_{n,\greekbf{\alpha}_n} ({\bf x},{\bf a}_t) &= \langle {\bf x} | \underline{n, \greekbf{\alpha}_{n}, {\bf a}_t }\rangle  \, ,\\
\overline{\rho}_{m,\greekbf{\beta}_m} ({\bf x},{\bf a}_t) &= \langle {\bf x} | \overline{m, \greekbf{\beta}_{m}, {\bf a}_t} \rangle  \,  .
\end{split}
\end{eqnarray}
Here we introduce the position basis $|{\bf x} \rangle$ which satisfies $(|{\bf x} \rangle)^\dagger = \langle {\bf x} |$ and 
\begin{eqnarray}
\begin{split}
\int \ud^D {\bf x} \, | {\bf x} \rangle \langle {\bf x} | &= 1 \, ,\\
\langle {\bf x} | {\bf x}^\prime \rangle &= \delta^{(D)} ({\bf x}-{\bf  x}^\prime) \, .
\end{split}
\end{eqnarray}
Then the conditions for the complete orthogonal set, Eq.\ (\ref{eqn:nor_ori}), 
are represented by 
\begin{eqnarray}
\begin{split}
\langle \overline{m, \greekbf{\beta}_{m}, {\bf a}_t} | \underline{n, \greekbf{\alpha}_{n} , {\bf a}_t} \rangle &= \delta_{m,n} \delta_{\greekbf{\beta}_{m}, \greekbf{\alpha}_{n}} \, ,\\
\sum_{n \ge 0} \sum_{\greekbf{\alpha}_{n} ={\bf 0}}^{{\bf d}^{(n)}({\bf a}_t)} 
| \underline{n, \greekbf{\alpha}_{n} , {\bf a}_t} \rangle \langle 
\overline{n, \greekbf{\alpha}_{n} , {\bf a}_t} | &= 1 \, ,
\end{split}
\end{eqnarray}
where
\begin{eqnarray}
\begin{split}
(| \overline{n, \greekbf{\alpha}_{n}, {\bf a}_t} \rangle)^\dagger &=& \langle \overline{n, \greekbf{\alpha}_{n}, {\bf a}_t} | \, ,\\
(| \underline{n, \greekbf{\alpha}_{n}, {\bf a}_t} \rangle)^\dagger &=& \langle \underline{n, \greekbf{\alpha}_{n}, {\bf a}_t} | \, .
\end{split}
\end{eqnarray}
From Eq.\ (\ref{eqn:ou1}), we find 
\begin{eqnarray}
 | \overline{n, \greekbf{\alpha}_{n} , {\bf a}_t} \rangle= 
\frac{1}{{\cal N}_{n,\alpha_n}({\bf a}_t)} e^{\widehat{G}({\bf a}_t)} | \underline{n, \greekbf{\alpha}_{n} , {\bf a}_t} \rangle
\, ,
\label{eqn:ou2} 
\end{eqnarray}
where the operator $\widehat{G}({\bf a}_t)$ is defined by
\begin{eqnarray}
\langle {\bf x} | \widehat{G} ({\bf a}_t) | {\bf x}^\prime \rangle = G({\bf x}, {\bf a}_t) \delta^{(D)} ({\bf x} - {\bf x}^\prime ) \, .
\end{eqnarray}

In this notation, the Fokker-Planck equation is represented by 
\begin{eqnarray}
\partial_t | \rho (t) \rangle = \widehat{\cal L}_t | \rho (t) \rangle \, ,
\end{eqnarray}
where  
\begin{eqnarray}
\begin{split}
\langle {\bf x} | \widehat{V}_t | {\bf x}^\prime \rangle &= V({\bf x},{\bf a}_t) \delta^{(D)} ({\bf x}-{\bf x}^\prime) \, , \\
\langle {\bf x} | \widehat{\cal L}_t | {\bf x}^\prime \rangle &= {\cal L}_t ({\bf x}) \delta^{(D)} ({\bf x}-{\bf x}^\prime)  \, .
\end{split}
\end{eqnarray}
The eigenvalue equations are then expressed as
\begin{eqnarray}
\begin{split}
\widehat{\cal L}_t | \underline{n, \greekbf{\alpha}_{n}, {\bf a}_t} \rangle &= - \lambda_n (t) | \underline{n, \greekbf{\alpha}_{n}, {\bf a}_t} \rangle \, , \label{eqn:eigen_l} \\
\widehat{\cal L}^\dagger_t  | \overline{m, \greekbf{\beta}_{m} ,{\bf a}_t} \rangle  &= - \lambda_m(t) | \overline{m, \greekbf{\beta}_{m} ,{\bf a}_t} \rangle \, .
\end{split}
\end{eqnarray}
Here we have used Eq.\ (\ref{eqn:lam=ov=un}).

One may notice that these representations in the bi-orthogonal basis look very similar to 
the non-Hermitian generalization of quantum mechanics
 (which is, sometimes, called $\mathcal{PT}$-symmetric quantum mechanics) \cite{bender,brody,most}. 
Indeed, the operator $e^{G({\bf x},{\bf a}_t)}/ {\cal N}_{n,\greekbf{\alpha}_n}({\bf a}_t)$ in Eq.\ (\ref{eqn:ou1}) 
seems to be a kind of the metric operator. Thus a head state is obtained from a foot state by multiplying it and vice versa. 
See also the discussion in Ref.\ \cite{koide18}.

\section{Perturbative solution of time-dependent Fokker-Planck equation} \label{sec:pertur-fp}

There are various expansions to express the solution of the Fokker-Planck equation. See Refs.\ 
\cite{talkner,dreger,hesam,welles2014,wadia} and references therein.
In our method, we expand the solution of the Fokker-Planck equation 
in terms of the foot states introduced in the previous section,
\begin{eqnarray}
| \rho (t) \rangle =  
\sum_{n \ge 0} \sum_{\greekbf{\alpha}_n ={\bf 0}}^{{\bf d}^{(n)}({\bf a}_t)}  A_{n,\greekbf{\alpha}_n} (t) 
e^{-\int^t_{t_i} \ud s \, \lambda_n (s)}
| \underline{n,\greekbf{\alpha}_n, {\bf a}_t} \rangle \, ,
\label{eqn:dec_rho}
\end{eqnarray}
where $t_i$ is an initial time.
Note that the sums in the above expansion can be time-dependent 
because of ${\bf d}^{(n)}({\bf a}_t)$.
In the following calculation, however, we consider the case where 
the sums associated with $n$ and $\greekbf{\alpha}_n$ are independent of time, 
\begin{eqnarray}
\sum_{n \ge 0} \sum_{\greekbf{\alpha}_n ={\bf 0}}^{{\bf d}^{(n)}({\bf a}_t)} 
=
\sum_{n \ge 0} \sum_{\greekbf{\alpha}_n = {\bf 0}}^{{\bf d}^{(n)}({\bf a}_{t^\prime})} \,  \,\,\,\,\, (t \neq t^\prime)\, .
\label{eqn:time-ind-sum}
\end{eqnarray}

For the sake of simplicity, 
we further decompose the coefficient $A_{n,\greekbf{\alpha}_n} (t)$ as 
\begin{eqnarray}
A_{n, \greekbf{\alpha}_n}(t) 
= 
\sum_{\greekbf{\beta}_n = {\bf 0}}^{{\bf d}^{(n)}({\bf a}_t) }[{\bf M}^{(n)}(t)]_{\greekbf{\alpha}_n,\greekbf{\beta}_n} 
C_{n,\greekbf{\beta}_n}(t) \, . \label{eqn:def_c}
\end{eqnarray}
Here the matrix $[{\bf M}^{(n)}(t)]_{\greekbf{\alpha}_n,\greekbf{\beta}_n} $ is obtained by the solution of the following differential equation:
\begin{eqnarray}
\partial_t [{\bf M}^{(n)}(t)]_{\greekbf{\alpha}_n,\greekbf{\beta}_n} 
= -\sum_{\greekbf{\gamma}_n ={\bf 0}}^{{\bf d}^{(n)}({\bf a}_t)} 
\langle \overline{n, \greekbf{\alpha}_n,{\bf a}_t} |\partial_t | \underline{n, \greekbf{\gamma}_n,{\bf a}_t} \rangle
[{\bf M}^{(n)}(t)]_{\greekbf{\gamma}_n,\greekbf{\beta}_n} 
\, ,
\end{eqnarray}  
with the initial condition 
\begin{eqnarray}
[{\bf M}^{(n)}(t_i)]_{\greekbf{\alpha}_n,\greekbf{\beta}_n} = \delta_{\greekbf{\alpha}_n, \, \greekbf{\beta}_n} \, .
\end{eqnarray}

Substituting this expansion into the Fokker-Planck equation, 
we find that the expansion coefficients are given by the solutions of the following differential equation:
\begin{eqnarray}
&& \partial_t C_{n,\greekbf{\alpha}_n}(t) 
 + 
\sum_{\greekbf{\beta}_n={\bf 0}}^{{\bf d}^{(n)}({\bf a}_t)} \sum_{m\neq n\ge 0} \sum_{\greekbf{\gamma}_m,\greekbf{\delta}_m={\bf 0}}^{{\bf d}^{(m)}({\bf a}_t)} 
e^{\int^t_{t_i} \ud s \, (\lambda_n(s) - \lambda_m(s))}  \nonumber \\
&& \times \frac{[{\bf M}^{(n)}(t)]^{-1}_{\greekbf{\alpha}_n,\greekbf{\beta}_n}\left[ \dot{\bf L}^{(n,m)} (t) \right]_{\greekbf{\beta}_n,\greekbf{\gamma}_m}
[{\bf M}^{(m)}(t)]_{\greekbf{\gamma}_m,\greekbf{\delta}_m} }{\lambda_n(t) - \lambda_m(t)}
C_{m,\greekbf{\delta}_m}(t)
 = 0 \, , \label{eqn:c}
\end{eqnarray}
where
\begin{eqnarray}
\left[ \dot{\bf L}^{(n,m)} (t) \right]_{\greekbf{\alpha}_n,\greekbf{\beta}_m} = 
\langle \overline{n, \greekbf{\alpha}_n, {\bf a}_t} | (\partial_t \widehat{\cal L}_t) | \underline{m, \greekbf{\beta}_m, {\bf a}_t} \rangle \, ,
\end{eqnarray}
and the inverse matrix is defined by
\begin{eqnarray}
\sum_{\greekbf{\gamma}_n ={\bf 0}}^{{\bf d}^{(n)}({\bf a}_t)} 
[{\bf M}^{(n)}(t)]^{-1}_{\greekbf{\alpha}_n,\greekbf{\gamma}_n}[{\bf M}^{(n)}(t)]_{\greekbf{\gamma}_n,\greekbf{\beta}_n} = \delta_{\greekbf{\alpha}_n,\, \greekbf{\beta}_n} \, .
\end{eqnarray}

It should be noted that the coefficient $C_{0,{\bf 0}}(t)$ is time-independent, 
\begin{eqnarray}
C_{0,{\bf 0}}(t) = C_{0,{\bf 0}}(t_i) \, , \label{eqn:cond_c00}
\end{eqnarray}
because
\begin{eqnarray}
\left[
\dot{\bf L}^{(0,m)} (t)
\right]_{{\bf 0},\greekbf{\beta}_m}
\propto 
\int 
\ud^D x \, 
\nabla \cdot \left\{
\underline{\rho}_{m,\greekbf{\beta}_m} ({\bf x},t)
 \nabla (\partial_t V ({\bf x}, {\bf a}_t) )
\right\} = 0\, .
\end{eqnarray}

\subsection{Evolution from equilibrium state}

In the following, we focus on the time evolution from a thermal equilibrium state with the control parameters ${\bf a}_{t_i}$,
\begin{eqnarray}
\rho_{eq} ({\bf x},t_i)  
= \frac{1}{{\cal Z}({\bf a}_{t_i})} e^{-\beta V({\bf x},{\bf a}_{t_i})} \, , \label{eqn:equili}
\end{eqnarray}
where ${\cal Z}({\bf a}_{t_i})$ is defined by Eq.\ (\ref{eqn:def_z}).
The expansion of this state in terms of the foot state leads to  
\begin{eqnarray}
\rho_{eq} ({\bf x},t_i)  =\sum_{n \ge 0} \sum_{\greekbf{\alpha}_n ={\bf 0}}^{{\bf d}^{(n)}({\bf a}_t)} 
C_{n,\greekbf{\alpha}_n} (t_i) \langle {\bf x} | \underline{n, \greekbf{\alpha_n} ,{\bf a}_{t_i}} \rangle \, ,
\end{eqnarray}
where
\begin{eqnarray}
\left\{
\begin{array}{ll}
C_{0,{\bf 0}} (t_i) = \frac{1}{\sqrt{{\cal N}_{0,{\bf 0}}(t_i)}} \frac{1}{\sqrt{{\cal Z} ({\bf a}_{t_i})}} 
= \langle \overline{0, {\bf 0}, {\bf a}_{t_i}} | {\bf x} \rangle & \\
C_{n, \greekbf{\alpha}_n} (t_i) = 0  &    (n > 0) 
\end{array}
\right.
\, .
\end{eqnarray}

For this initial condition, Eq.\ (\ref{eqn:c}) is further simplified. 
Let us reexpress the expansion coefficient $C_{n,\greekbf{\alpha}_n} (t)$ as 
\begin{eqnarray}
C_{n,\greekbf{\alpha}_n} (t) =\langle \overline{0, {\bf 0}, {\bf a}_{t_i}} | {\bf x} \rangle D_{n,\greekbf{\alpha_n}} (t) \, .
\end{eqnarray}
Then, by solving Eq.\ (\ref{eqn:c}) formally, the expansion coefficient $D_{n,\greekbf{\alpha_n}} (t)$ 
is determined by solving the following equation iteratively:
\begin{eqnarray}
D_{n, \greekbf{\alpha}_n}(t)
&=& 
D_{n,\greekbf{\alpha}_n}(t_i)
- 
\int^{t}_{t_i} \ud s \,
\sum_{\greekbf{\beta}_n={\bf 0}}^{{\bf d}^{(n)}({\bf a}_s)} 
\sum_{m\neq n \ge 0} \sum_{\greekbf{\gamma}_m,\greekbf{\delta}_m={\bf 0}}^{{\bf d}^{(m)}({\bf a}_s)} 
e^{\int^s_{t_i} \ud s^\prime \, (\lambda_n(s^\prime) - \lambda_m(s^\prime))}  \nonumber \\
&& \times 
\frac{[{\bf M}^{(n)}(s)]^{-1}_{\greekbf{\alpha}_n,\greekbf{\beta}_n}\left[ \dot{\bf L}^{(n,m)} (s) \right]
_{\greekbf{\beta}_n,\greekbf{\gamma}_m}
[{\bf M}^{(m)}(s)]_{\greekbf{\gamma}_m,\greekbf{\delta}_m}}{\lambda_n(s) - \lambda_m(s)} 
D_{m,\greekbf{\delta}_m}(s)\, , 
\label{eqn:d}
\end{eqnarray}
with the initial conditions 
\begin{eqnarray}
D_{n,\greekbf{\alpha}_n} (t_i) = \delta_{n,0} \delta_{\greekbf{\alpha}_n,\, {\bf 0}} \label{eqn:ini_cond_D} \, .
\end{eqnarray}
Again, note that $D_{0,{\bf 0}} (t)$ is independent of time and thus 
\begin{eqnarray}
D_{0,{\bf 0}} (t) = D_{0,{\bf 0}} (t_i) \, ,
\end{eqnarray} 
because of Eq.\ (\ref{eqn:cond_c00}).
In the following calculations, we use exclusively this expansion coefficient $D_{n,\greekbf{\alpha_n}} (t)$.

\subsection{Pseudo density matrix}

Let us introduce the pseudo density matrix defined by 
\begin{eqnarray}
\widehat{\rho}(t)
= 
\sum_{n\ge 0} \sum_{\greekbf{\alpha}_n,\greekbf{\beta}_n={\bf 0}}^{{\bf d}^{(n)}(t)} e^{-\int^t_{t_i} \ud s \, \lambda_n (s) }
| \underline{n, \greekbf{\alpha}_n,{\bf a}_t} \rangle 
[{\bf M}^{(n)}(t)]_{\greekbf{\alpha}_n,\greekbf{\beta}_n} D_{n,\greekbf{\beta}_n}(t)
\langle \overline{0, {\bf 0},{\bf a}_{t_i}} | \, . \label{eqn:pdm}
\end{eqnarray}
This satisfies the following properties: 
\begin{eqnarray}
{\rm Tr} [\widehat{\rho} (t)] &=& 1 \, , \\
\widehat{\rho}^2(t) &=& \widehat{\rho} (t) \, .
\end{eqnarray}
To show the second equation, we have used
\begin{eqnarray}
\langle \overline{0,{\bf 0},{\bf a}_{t_i}}|  = [{\bf M}^{(0)}(t)]^{-1}_{{\bf 0},{\bf 0}}  \langle \overline{0,{\bf 0},{\bf a}_t} | \, .
\label{eqn:formula1}
\end{eqnarray}

The expectation values are represented 
by the trace with the pseudo density matrix. 
For example, the mean work induced by the deformation of the external potential is given by 
\begin{eqnarray}
W = \int \ud t\,  {\rm Tr} [ (\partial_t \widehat{V}_t) 
\widehat{\rho}(t) ] \, .
\end{eqnarray}

\section{Formula for mean work} \label{sec:formula_mw}

Let us consider the time evolution from a thermal equilibrium state which is induced 
by changing the control parameters ${\bf a}_t$ in a finite time interval $t_i \le t \le t_f$ where $t_f$ is a finial time.
The mean work induced by the deformation of the potential is given by 
the time integral of Eq.\ (\ref{eqn:meanwork}). Using Eq.\ (\ref{eqn:pdm}), we find 
\begin{eqnarray}
W 
&=&
\int^{t_f}_{t_i} \ud t\, \ud W \nonumber \\
%&=&
%\int^{t_f}_{t_i} \ud t\,  {\rm Tr} [ (\partial_t \widehat{V}) \widehat{\rho}(t) ] \nonumber \\
&=& 
 \int^{t_f}_{t_i} \ud t\,  
\sum_{n\ge 0} \sum_{\greekbf{\alpha}_n,\greekbf{\beta}_n={\bf 0}}^{{\bf d}^{(n)} ({\bf a}_t)} e^{-\int^t_{t_i}\ud \,  \lambda_n (s) }
[{\bf M}^{(n)}(t)]_{\greekbf{\alpha}_n,\greekbf{\beta}_n} D_{n,\greekbf{\beta}_n}(t) 
\langle \overline{0, {\bf 0}, {\bf a}_{t_i}} |(\partial_t \widehat{V}_t)| \underline{n,\greekbf{\alpha}_n,{\bf a}_t} \rangle \, .
\label{eqn:def_work}
\end{eqnarray}
This formula is the generalization of Eq.\ (38) of Ref. \cite{koide17} to the systems in higher spatial dimensions.
Indeed, this is reduced to Eq.\ (38) when we set 
$[{\bf M}^{(n)}(t)]_{\greekbf{\alpha}_n,\greekbf{\beta}_n} = e^{-\theta_n (t)} \delta_{\greekbf{\alpha}_n,{\bf 0}}\delta_{\greekbf{\beta}_n,{\bf 0}}$.
The role of $\theta_n (t)$ is discussed in Sec.\ \ref{sec:berry}.

To expand the right-hand side in terms of the time derivative of the control parameters, $\dot{\bf a}_t$,
we introduce the expansions of the mean work $W$ and the coefficients 
$D_{n,\greekbf{\beta}_n}(t)$, 
\begin{eqnarray}
W &=& W^{(1)} + W^{(2)} + \cdots \, , \\
D_{n,\greekbf{\beta}_n}(t) 
&=& D^{(0)}_{n,\greekbf{\beta}_n}(t) + D^{(1)}_{n,\greekbf{\beta}_n}(t) + \cdots \, ,
\end{eqnarray}
where $W^{(m)}$ and $D^{(m)}_{n,\greekbf{\beta}_n}(t)$ indicate the $m$-th order terms of $\dot{\bf a}_t$.
From the initial condition (\ref{eqn:ini_cond_D}), we find 
\begin{eqnarray}
\left\{
\begin{array}{ll}
D^{(0)}_{n,\greekbf{\beta}_n}(t) = \delta_{n,0} \delta_{\greekbf{\beta}_n, {\bf 0}} &  \\
D^{(m)}_{0,{\bf 0}}(t)  = 0  & (m\ge 1)
\end{array}
\, . \right.
\label{eqn:condini_d}
\end{eqnarray}

\subsection{Lowest order contribution $W^{(1)}$}

The lowest order contribution in the mean work is given by $W^{(1)}$, 
\begin{eqnarray}
W^{(1)} 
=  
\int^{t_f}_{t_i} \ud t\,  
[{\bf M}^{(0)}(t)]_{{\bf 0},{\bf 0}} 
\langle \overline{0, {\bf 0}, {\bf a}_{t_i}} |(\partial_t \widehat{V}_t)| \underline{0,{\bf 0},{\bf a}_t} \rangle \, .
\label{eqn:w1}
\end{eqnarray}
This quantity is calculated exactly without referring to the detailed behavior of the control parameters ${\bf a}_t$, and 
given by the difference of the free energies, 
\begin{eqnarray}
W^{(1)}
=
F ({\bf a}_{t_f}) - F ({\bf a}_{t_i})
\, , \label{eqn:re_work}
\end{eqnarray}
where
\begin{eqnarray}
F({\bf a}_t) = -\frac{1}{\beta} \ln {\cal Z} ({\bf a}_t) 
= - \frac{1}{\beta} \ln \left( \int  \ud^D x\,  e^{-\beta V({\bf x},{\bf a}_t)} \right) 
\label{def:fmesmo}
\, . 
\end{eqnarray}
It should be emphasized that this free energy is not the Helmholtz free energy defined by Eq.\ (\ref{eqn:helmholtz}).
We see that $W^{(1)}$ is determined only by the initial and final values of ${\bf a}_t$ 
and independent of the intermediate deformation processes of $V({\bf x},{\bf a}_t)$.

By definition, the initial state is given by a thermal equilibrium state and thus this free energy agrees with the 
Helmholtz free energy (\ref{eqn:helmholtz}) at the initial time $t_i$,
$F({\bf a}_{t_i}) = F_{helm} ({\bf a}_{t_i})$.
In general, $F({\bf a}_t)$ for $t > t_i$ does not necessarily agree with $F_{helm} ({\bf a}_{t})$. 
However, in the quasi-static process, the final state is approximately given by a thermal equilibrium state and then 
the right-hand side of Eq.\ (\ref{eqn:re_work}) is identified with the change of the Helmholtz free energy. 
This is the well-known behavior in the reversible process and thus $W^{(1)}$ is identified with the reversible work in this limit.

\subsection{Higher order contributions}

All higher order terms in the mean work  $W^{(n)}$ ($n\ge 2$) give irreversible contributions.
The term of $n=2$, $W^{(2)}$, is obtained  
by using $D^{(1)}_{n,\greekbf{\beta}_n}(t)$,
\begin{eqnarray}
W^{(2)} 
&=& 
\int^{t_f}_{t_i} \ud t\,  
\sum_{n\ge 1} \sum_{\greekbf{\alpha}_n,\greekbf{\beta}_n=0}^{{\bf d}^{(n)}({\bf a}_t)} e^{-\int^t_{t_i}\ud s\,  \lambda_n (s) }
[{\bf M}^{(n)}(t)]_{\greekbf{\alpha}_n,\greekbf{\beta}_n} D^{(1)}_{n,\greekbf{\beta}_n}(t) 
\langle \overline{0,0, {\bf a}_{t_i}} |(\partial_t \widehat{V}_t)| \underline{n,\greekbf{\alpha}_n,{\bf a}_t} \rangle \, ,
\label{eqn:exact_w2}
\end{eqnarray}
where
\begin{eqnarray}
D^{(1)}_{n,\greekbf{\beta}_n}(t) 
=
- (1-\delta_{n,0})
\int^{t}_{t_i} \ud s \, 
\sum_{\greekbf{\gamma}_n=0}^{{\bf d}^{(n)}({\bf a}_s)}  
e^{\int^s_{t_i} \ud s^\prime \, \lambda_n(s^\prime)} 
\frac{[{\bf M}^{(n)}(s)]^{-1}_{\greekbf{\beta}_n,\greekbf{\gamma}_n}\left[ \dot{\bf L}^{(n,0)} (s) \right]_{\greekbf{\gamma}_n,{\bf 0}}
[{\bf M}^{(0)}(s)]
}{\lambda_n(s)} \, .
\label{eqn:dn1}
\end{eqnarray}
The factor $ (1-\delta_{n,0})$ is due to the property (\ref{eqn:ini_cond_D}).
When $\widehat{V}_t$ is given by the harmonic potential, 
higher order terms $W^{(n)}$ $(n \ge 3)$ vanish and the exact mean work is given by 
\begin{eqnarray}
W = W^{(1)} + W^{(2)} \, .
\end{eqnarray}
This is discussed in detail soon later.

\subsection{Relation to Berry's phase} \label{sec:berry}

In quantum mechanics, it is known that the wave function acquires a phase induced by a cyclic adiabatic motion of a system.
This additional phase is affected by the geometrical properties of the parameter of the Hamiltonian 
and called Berry's geometric phase.

In the quasi-static limit, the solution of the Fokker-Planck equation is obtained by substituting $D_{n,\greekbf{\alpha_n}} (t)=D^{(1)}_{n,\greekbf{\alpha_n}} (t)$, 
\begin{eqnarray}
| \rho (t) \rangle 
=
\langle \overline{0, {\bf 0}, {\bf a}_{t_i}} | {\bf x} \rangle
e^{-\theta_0 (t)}
| \underline{0,{\bf 0}, {\bf a}_t} \rangle \, ,
\end{eqnarray}
where
\begin{eqnarray}
\theta_0 (t)= 
- \ln [{\bf M}^{(0)}(t)]_{{\bf 0},{\bf 0}}
=\int^t_{t_i} \ud s \langle \overline{0,{\bf 0}, {\bf a}_s} | \partial_s | \underline{0,{\bf 0}, {\bf a}_s} \rangle  \,  .
\end{eqnarray}
We observe that a non-trivial time-dependent factor $e^{-\theta_0 (t)}$ appears besides the time evolution of $| \underline{0,{\bf 0}, {\bf a}_t} \rangle $ 
and this factor corresponds to Berry's geometric phase.
See also the discussion in Ref.\ \cite{koide17}.

Wilczek and Zee found that Berry's phase becomes a matrix in the system with degeneracy \cite{wilczek-zee}. 
Indeed, $[{\bf M}^{(n)}(t)]_{\greekbf{\alpha}_n,\greekbf{\beta}_n}$ is generally given by a matrix for $n \neq 0$.
In the present calculation, the ground foot state is always not degenerate.  
Therefore, to observe the Wilczek-Zee-type phase, we should consider the time evolutions in the ``excited" state 
of the Fokker-Planck operator.  

\section{Application (I) : symmetric process} \label{sec:sym_prot}

To see the applications of the derived formula, 
we consider the two-dimensional harmonic potential,  
\begin{eqnarray}
V({\bf x},{\bf a}_t) = \frac{1}{2}a^{(1)}_t x^2_1 + \frac{1}{2}a^{(2)}_t x^2_2 \, . \label{eqn:harmonic_pot}
\end{eqnarray}
The Fokker-Planck operator with this potential is  expressed as
\begin{eqnarray}
{\cal L}_t ({\bf x}) 
&=& 
- e^{-G({\bf x},{\bf a}_t)} \left\{{\cal H}^{(1)}_t (x_1)+ {\cal H}^{(2)}_t (x_2)\right\} e^{G({\bf x},{\bf a}_t)}
\, ,
\end{eqnarray}
where
\begin{eqnarray}
{\cal H}^{(i)}_t (x_i)
&=& 
-\frac{1}{\nu\beta} \partial^2_i - \frac{a^{(i)}_t}{2\nu} 
+\frac{\beta}{4\nu} (a^{(i)}_t x_i)^2 \, , \\
G({\bf x},{\bf a}_t)
&=& 
\sum_{i=1}^2
\frac{\beta}{2} a^{(i)}_t x^2_i  \, .
\end{eqnarray}
To construct the foot and head states, we introduce the eigenfunctions of ${\cal H}^{(i)}_t(x_i)$ which are given by
\begin{eqnarray}
u_{n} (x_i,a^{(i)}_t) 
&=& 
\frac{1}{\sqrt{Z(a^{(i)}_t)}}\sqrt{\frac{1}{2^n n!} } e^{-\beta a^{(i)}_t x^2_i/4}
H_n (\sqrt{\beta a^{(i)}_t/2} x_i) \, ,  \label{eqn:def_u^i}
\end{eqnarray}
where $H_n (x)$ denote the Hermite polynomials and 
\begin{eqnarray}
Z(a) = \sqrt{\frac{2\pi}{\beta a}} \, .
\end{eqnarray}

In the two-dimensional system, the degeneracy related to the rotational symmetry of the potential is characterized by a single integer.
Let us consider the deformation of the harmonic potential with $a^{(1)}_t = a^{(2)}_t = a_t$.
Then the potential has the rotational symmetry around $(x_1,x_2) = (0,0)$ in the two-dimensional plane during the deformation process.
Then $\ud^{(n)}(a_t)$ are given by time-independent constants, $\ud^{(n)}$, and thus 
the condition (\ref{eqn:time-ind-sum}) is satisfied.
The foot and head states are defined by 
\begin{eqnarray}
\langle {\bf x} | \underline{n,\alpha_n, {\bf a}_t \rangle} 
&=& 
\underline{\rho}_{\alpha_n} (x_1,a_t)\underline{\rho}_{n-\alpha_n} (x_2,a_t) \, ,\\
\langle {\bf x} | \overline{n,\alpha_n, {\bf a}_t \rangle} 
&=& 
\overline{\rho}_{\alpha_n} (x_1,a_t)\overline{\rho}_{n-\alpha_n} (x_2,a_t) \, ,
\end{eqnarray}
where $\alpha_n$ take non-negative integers satisfying $0 \le \alpha_n \le \ud^{(n)} = n$, and  
\begin{eqnarray}
\underline{\rho}_{n} (x,a_t)
&=& 
\sqrt{N_{n}(a_t)}
e^{-\beta a_t x^2/4}
u_n (x,a_t)
\, , \label{eqn:ho_foot}\\
\overline{\rho}_{n} (x,a_t)
&=& 
\frac{1}{\sqrt{N_{n}(a_t)}}
e^{\beta a_t x^2/2}
u_n (x,a_t)
\, . \label{eqn:ho_head}
\end{eqnarray}
Here $N_{n}(a_t)$ is an arbitrary real function associated with the ambiguity of the normalization of the foot states discussed below Eq.\ (\ref{eqn:urho_u}). 
The eigenvalues of $| \underline{n,\alpha_n, {\bf a}_t \rangle}$ (and equivalently of $ | \overline{n,\alpha_n, {\bf a}_t \rangle} $) are given by
\begin{eqnarray}
\lambda_n (t)
= \frac{a_t}{\nu} n \,\,\,\,\,\,  (n\ge 0) \, .
\end{eqnarray}

The lowest order contribution in the mean work $W^{(1)}$ is already calculated. 
The next order contribution is given by Eq.\ (\ref{eqn:exact_w2}), which is simplified as 
\begin{eqnarray}
 W^{(2)}_{sym}
=
\int^{t_f}_{t_i} \ud t\,  
 \sum_{\alpha_2,\beta_2=0}^{\ud^{(2)}} e^{-\int^t_{t_i}\ud \,  \lambda_2 (s) }
[{\bf M}^{(2)}(t)]_{\alpha_2,\beta_2} D^{(1)}_{2,\beta_2}(t) 
\langle \overline{0, 0, {\bf a}_{t_i}} |(\partial_t \widehat{V}_t)| \underline{2, \alpha_2,{\bf a}_t} \rangle \, ,
\label{eqn:w2sym_still}
\end{eqnarray}
because, in the harmonic potential, we find
\begin{eqnarray}
\lefteqn{\langle \overline{0,0, {\bf a}_\tau} |  (\partial_t \widehat{V}) | \underline{n,\alpha_n, {\bf a}_t} \rangle} && 
\nonumber \\
&=& 
\frac{1}{2}\frac{\ud a_t}{\ud t}
%\sqrt{\frac{{\cal N}_{0,0}(t)}{{\cal N}_{0,0}(\tau)}} 
\sqrt{\frac{N^2_{0}(a_t)}{N^2_{0}(a_\tau)}} 
%e^{2\ii (\delta_0 (a_t) - \delta_0(a_{t_i}) )}
\left( \frac{a_\tau}{a_t} \right)^{1/2} \nonumber \\
&& \times \frac{1}{\beta a_t}
\left[
 \delta_{n,0} e^{-\ii (\delta_{n+2} (a_t) - \delta_n (a_t))}
+ \sqrt{\frac{N_2(a_t)}{N_{0}(a_t)}} 
e^{-\ii (\delta_{n-2} (a_t) - \delta_n (a_t))}
\sqrt{2} 
 \delta_{n,2} 
\right] 
\left[
\delta_{0,\alpha_n} + \delta_{n,\alpha_n}
\right] \, . \nonumber \\
\end{eqnarray}
The expansion coefficient $D^{(1)}_{2,\beta_2}(t)$ is given by Eq.\ (\ref{eqn:dn1}).
Substituting it into Eq.\ (\ref{eqn:w2sym_still}), we obtain
\begin{eqnarray}
W^{(2)}_{sym}
&=&
2\int^{t_f}_{t_i} \ud t \int^{t}_{t_i} \ud s \, 
\frac{\ud a_s}{\ud s}
\left[ \frac{1}{2\beta a^2_s} e^{-\int^t_{s}\ud s^\prime \frac{2a_{s^\prime}}{\nu} }
\right]
\frac{\ud a_t}{\ud t} \, , \label{eqn:exa_irw_2d}
\end{eqnarray}
where we have used Eq.\ (\ref{eqn:formula1}) and 
the following results:
\begin{eqnarray}
[{\bf M}^{(n)}(t) ]_{\alpha_n,\beta_n} 
&=& 
\delta_{\alpha_n,\beta_n} 
\sqrt{ \frac{N_{\alpha_n} (a_{t_i}) N_{n-\alpha_n} (a_{t_i})}{N_{\alpha_n} (a_t) N_{n-\alpha_n} (a_t) }}
\left( \frac{a_t}{a_{t_i}} \right)^{(n+1)/2} 
\, , \nonumber \\
\\
\left[ \dot{\bf L}^{(n,m)} (\tau) \right]_{\alpha_n,\beta_m} 
&=& 
-\frac{1}{\nu} \frac{\ud a_\tau}{\ud \tau} \left[
 m  \delta_{n,m} \delta_{\alpha_n,\beta_m} 
 + \sqrt{\frac{N_{\beta_m}(a_\tau)}{N_{\beta_m + 2}(a_\tau)}}
\sqrt{(\beta_m + 2)(\beta_m + 1)} \delta_{n,m+2} \delta_{\alpha_n,\beta_m+2} \right. \nonumber \\
&& \left. + \sqrt{\frac{N_{m-\beta_m}(a_\tau)}{N_{m-\beta_m + 2}(a_\tau)}}
\sqrt{(m-\beta_m + 2)(m-\beta_m + 1)}  \delta_{n,m+2} \delta_{\alpha_n,\beta_m} 
\right]\, .
\label{eqn:dotL}
\end{eqnarray}

In the harmonic potential, 
the exact mean work is given by 
\begin{eqnarray}
W _{sym}= W^{(1)} + W^{(2)}_{sym} \, . \label{eqn:worksum}
\end{eqnarray} 
To understand this, note that, as seen from Eqs.\ (\ref{eqn:d}) and (\ref{eqn:condini_d}), 
the expansion coefficient $D^{(n)}_{2,\beta_2}(t)$ has the following product:
\begin{eqnarray}
&& \sum_{m_1 \neq 2} \sum_{m_2 \neq m_1} \cdots \sum_{m_{n-1} \neq 0} 
[{\bf M}^{(2)} (\tau_1)]^{-1}  \left[ \dot{\bf L}^{(2,m_1)} (\tau_1) \right] [{\bf M}^{(m_1)}(\tau_1)] \nonumber \\
&& \times [{\bf M}^{(m_1)} (\tau_2)]^{-1}  \left[ \dot{\bf L}^{(m_1,m_2)} (\tau_2) \right] [{\bf M}^{(m_2)}(\tau_2)] 
\cdots  \nonumber \\
&& \times 
[{\bf M}^{(m_{n-1})} (\tau_n)]^{-1} \left[ \dot{\bf L}^{(m_{n-1},0)} (\tau_n) \right] [{\bf M}^{(0)}(\tau_n)] \, .
\label{eqn:zeo_higherorder}
\end{eqnarray}
From Eq.\ (\ref{eqn:dotL}), it is easy to see that 
$\left[ \dot{\bf L}^{(n,m)} (\tau) \right]$ has a finite contribution only when 
$n=m+2$ and $n=m$, but the latter is excluded in the sums.
Therefore all higher order terms $D^{(n)}_{2,\beta_2}(t)$ ($n\ge 2$) disappear.

The mean work in the one-dimensional harmonic potential is obtained in Ref.\ \cite{koide17}.
Comparing the one-dimensional and two-dimensional results, 
we find that $W^{(2)}_{sym}$ in the two-dimensional calculation is twice as large as the one-dimensional result given by Eq.\ (50) in Ref.\ \cite{koide17}. 
The same property is found in the three-dimensional symmetric deformation of the harmonic potential.
Therefore the D-dimensional result is comprehensively summarized by  
\begin{eqnarray}
W^{(2)}_{sym}
= 
D \int^{t_f}_{t_i} \ud t \int^{t}_{t_i} \ud s \, 
\frac{\ud a_s}{\ud s}
\left[ \frac{1}{2\beta a^2_s} e^{-\int^t_{s} \ud s^\prime \frac{2a_{s^\prime}}{\nu} }
\right]
\frac{\ud a_t}{\ud t}
 \, , \label{eqn:sym_w2_d}
\end{eqnarray}
where $D$ is the number of the spatial dimension and can take $D=1,2$ or $3$. 
This simple $D$ dependence is understood by the fact that 
the harmonic potential is given by the independent sum of the each spatial component 
and thus the Fokker-Planck operator is represented by the independent sum of the spatial components, 
\begin{eqnarray}
{\cal L}_t = \sum_{i=1}^D \left[  \frac{1}{\nu \beta} \partial^2_i - \partial_i \frac{a_t x_i}{\nu}  \right] \, .
\end{eqnarray}
Therefore 
the optimization with respect to the $i$-th spatial component does not affect that to the $j$-th spatial components when $i\neq j$, 
and the mean work is given by a linear function of $D$ in the present symmetric process.

\subsection{Energy and Thermodynamic Relation} \label{sec:ener}

Following the thermodynamic interpretation introduced in Sec.\ \ref{sec:se}, 
the mean energy of this system is defined by Eq.\ (\ref{eqn:def_ener}), 
\begin{eqnarray}
E(t) = {\rm Tr}[ \widehat{V}_t \widehat{\rho}(t) ] \, .
\end{eqnarray}
Applying the same procedure used in the calculation of the mean work, 
the exact mean energy is calculated by using $D^{(1)}_{2,\beta_2}$ in the pseudo density matrix.
The result is given by 
\begin{eqnarray}
E_{sym}(t)
= 
\frac{1}{\beta} 
+ \frac{a_t}{\beta} \int^t_{t_i} \ud s\, e^{-\int^t_s \ud s^\prime \frac{2a_{s^\prime}}{\nu}} \frac{1}{a^2_s} 
\frac{\ud a_s}{\ud s}
\, .
\end{eqnarray}
Because $E_{sym}(t_i) =1/\beta$,  
the change of the mean energy induced in the symmetric process is given by the second term of the above equation, 
\begin{eqnarray}
\Delta E_{sym} = E_{sym}(t_f) - E_{sym}(t_i)
 =  \frac{a_{t_f}}{\beta} \int^{t_f}_{t_i} \ud s\, 
e^{-\int^{t_f}_s \ud s^\prime\, \frac{2a_{s^\prime}}{\nu}} \frac{1}{a^2_s} 
\frac{\ud a_s}{\ud s} \, .
\label{eqn:deltae_sym}
\end{eqnarray}

\subsection{Optimization in symmetric process} \label{sec:opt}

We are in particular interested in the optimized process which minimizes the mean work.
Because $W^{(1)}$ is determined only by ${\bf a}_{t_i}$ and ${\bf a}_{t_f}$ independently of processes, 
it is sufficient to consider the variation of $W^{(2)}_{sym}$ by fixing the operation time $\tau_{op}= t_f - t_i$.
The variation of the control parameter is defined by 
\begin{eqnarray}
a_t \longrightarrow a_t + \delta a_t \, .
\end{eqnarray} 
We fix the initial and final forms of the harmonic potential and hence this variation satisfies the boundary conditions, 
\begin{eqnarray}
\delta a_{t_i} = \delta a_{t_f} = 0 \, .
\end{eqnarray}
The variation of $W^{(2)}_{sym}$ leads to 
\begin{eqnarray}
&& \int^t_{t_i} \ud s\, \frac{1}{a^2_s} \frac{\ud a_s}{\ud s} \partial_t e^{-\frac{2}{\nu}\int^t_s \ud \tau\, a_\tau}
+ \frac{1}{a^2_t} \int^{t_f}_t \ud s\, \frac{\ud a_s}{\ud s} \partial_t e^{-\frac{2}{\nu} \int^s_t \ud \tau\, a_\tau} 
\nonumber \\
&& + \frac{2}{\nu} \int^{t_f}_t \ud s_2 \int^t_{t_i} \ud s_1\, e^{-\frac{2}{\nu}\int^{s_2}_{s_1}\ud \tau\, a_\tau }
\frac{1}{a^2_{s_1}}\frac{\ud a_{s_1}}{\ud s_1}\frac{\ud a_{s_2}}{\ud s_2} = 0 \, .
\label{eqn:diff_opt_sol}
\end{eqnarray}
Because of the reason discussed below Eq.\ (\ref{eqn:sym_w2_d}), 
this integro-differential equation is the same as Eq.\ (59) of Ref. \cite{koide17} which determines the optimized process in the one-dimensional harmonic potential.
See Appendix B in Ref. \cite{koide17} for the detailed derivation.

The optimized parameter is determined by solving this equation, but, so far, 
the exact solution is not known.
Therefore we consider an approximated process in the large $\tau_{op}$ limit as is discussed in Ref.\ \cite{koide17}.
For this purpose, we reexpress $W^{(2)}_{sym}$ in the adimensional form,
\begin{eqnarray}
W^{(2)}_{sym}
&=& 
2 \int^{1}_{0} \ud \tau_1 \int^{\tau_1}_{0} \ud \tau_2 \, 
\frac{\ud \bar{a}_{\tau_2}}{\ud \tau_2}
\left[ 
 e^{-\frac{2\tau_{op} a_{t_i}}{\nu} \int^{\tau_1}_{\tau_2}d\tau_3 \, \bar{a}_{\tau_3} }
\frac{1}{2 \beta (\bar{a}_{\tau_2})^2} 
\right]
\frac{\ud \bar{a}_{\tau_1}}{\ud \tau_1} \, ,
\label{eqn:wirr_sym_adim}
\end{eqnarray}
where adimensional quantities are introduced by 
\begin{eqnarray}
\tau &=& (t-t_i)/\tau_{op} \, , \\
\bar{a}_\tau &=& \frac{a_{t_i + \tau \tau_{op}}}{a_{t_i}} \, .
\end{eqnarray}
In the large $\tau_{op}$ limit, we consider the following approximation:  
\begin{eqnarray}
\int^{1}_0 \ud \tau_1 \int^{\tau_1}_0 \ud \tau_2 \, G(\tau_1,\tau_2) e^{-\frac{2\tau_{op} a_{t_i}}{\nu} \int^{\tau_1}_{\tau_2}d\tau_3 \, \bar{a}_{\tau_3} }
&\approx&
\int^{1}_0 \ud \tau_1 \int^{\tau_1}_0 \ud \tau_2 \, G(\tau_1,\tau_1) 
e^{-\frac{2\tau_{op} a_{t_i}}{\nu} \bar{a}_{\tau_1} (\tau_1 - \tau_2) } \nonumber \\
&\approx& 
\int^{1}_0 \ud \tau_1\, G(\tau_1,\tau_1) \frac{\nu}{2\tau_{op} a_i \bar{a}_{\tau_1} }  \, ,
\end{eqnarray}
where $G(\tau_1,\tau_2)$ is a smooth function.
Using this approximation, $W^{(2)}_{sym}$ is calculated as  
\begin{eqnarray}
W^{(2)}_{sym} = \frac{\nu}{2\beta a_{t_i} \tau_{op}}\int^1_0 \ud \tau \,  
\frac{1}{(\bar{a}_{\tau})^3} 
\left( \frac{\ud \bar{a}_{\tau}}{\ud \tau} \right)^2 \, .
\label{eqn:app_irrw_sym}
\end{eqnarray}
We minimize this approximated mean work.
Applying the variation defined above to this, we find
\begin{eqnarray}
\frac{1}{\bar{a}^3_{\tau}} \frac{\ud^2 \bar{a}_\tau }{\ud \tau^2}- \frac{3}{2} \frac{1}{\bar{a}^4_\tau} 
\left( \frac{\ud \bar{a}_\tau}{\ud \tau}\right)^2= 0 \, . \label{eqn:opt_a}
\end{eqnarray}
This equation is analytically solvable and the optimized external parameter in the large $\tau_{op}$ limit is given by 
\begin{eqnarray}
\bar{a}_\tau
= 
\frac{\bar{a}_1}{ \left[ \sqrt{\bar{a}_1}- (\sqrt{\bar{a}_1} - 1)\tau \right]^2}
\longrightarrow
a_t = \frac{\tau_{op}^2 \, a_{t_i} a_{t_f}}{\left[ (t-t_i) \sqrt{a_{t_i}} + (t_f - t) \sqrt{a_{t_f}} \right]^2} \, .
\label{eqn:app_protocol} 
\end{eqnarray}
This is the same as those found in one-dimensional harmonic oscillator \cite{sekimoto-sasa,schmiedl-seifert2007,koide17}.

Substituting this into Eq.\ (\ref{eqn:app_irrw_sym}), 
the optimized mean work in the large $\tau_{op}$ limit is expressed as
\begin{eqnarray}
\lim_{\tau_{op} \rightarrow \infty} \frac{W_{sym}}{W^{(1)}} 
= 
\lim_{\tau_{op} \rightarrow \infty} \frac{W^{(1)} + W^{(2)}_{sym}}{W^{(1)}} 
= 
1+ 
\frac{2}{\bar{a}_1 \ln \bar{a}_1}  \left( \sqrt{\bar{a}_1} - 1 \right)^2
\frac{\nu}{a_{t_i} \tau_{op}} \, ,
\label{eqn:irr_w_sym}
\end{eqnarray}
where, from Eq.\ (\ref{eqn:re_work}),  
\begin{eqnarray}
W^{(1)} = -\frac{1}{\beta} \ln \frac{{\cal Z}(a_{t_f})}{{\cal Z}(a_{t_i})} = \frac{1}{\beta} \ln \bar{a}_1  \, . \label{eqn:w(1)}
\end{eqnarray}
The time evolution in the large $\tau_{op}$ limit is very slow 
and thus $\rho({\bf x},t_f)$ is approximately given by the thermal equilibrium state with the parameter $a_{t_f}$, $\rho_{eq} ({\bf x},t_f)$. 
Then $W^{(1)}$ in Eq.\ (\ref{eqn:re_work}) corresponds to the one in the reversible process 
and is represented by the change of the Helmholtz free energy.
In this limit, one can see that the irreversible contribution of the mean work asymptotically disappears 
as a function of  $\tau^{-1}_{op}$ in Eq.\ (\ref{eqn:irr_w_sym}). 
It is worth emphasizing that this $\tau_{op}$ dependence is experimentally verified \cite{blickle2}.

In a similar fashion, the change of the mean energy (\ref{eqn:deltae_sym}) is calculated with this approximated optimized process (\ref{eqn:app_protocol}). 
In the large $\tau_{op}$ limit, this quantity is simplified as 
\begin{eqnarray}
\Delta E_{sym} \approx \frac{\nu}{2\tau_{op} a_{t_i}\beta} \frac{1}{\bar{a}^2_1} 
\left. \frac{\ud \bar{a}_\tau}{\ud \tau} \right|_{\tau=1} \, .
\end{eqnarray}
Substituting Eq.\ (\ref{eqn:app_protocol}) into this, the change of the mean energy is calculated as
\begin{eqnarray}
\lim_{\tau_{op} \rightarrow \infty} \frac{ \Delta E_{sym}}{E(t_i)} 
= \frac{1}{\bar{a_1} } \left( \sqrt{\bar{a}_1} - 1 \right)  \frac{\nu}{a_{t_i} \tau_{op}}
\, .
\label{eqn:mean_e_sym_large}
\end{eqnarray}
This equation implies that there is no change of the mean energy in the quasi-static process.
This is reminiscent of the quasi-static isothermal process in the ideal gas.

Later, we will discuss the asymmetric deformation of the harmonic potential 
and compare the results with those in the symmetric process. 
In the asymmetric process, 
we choose the initial and final parameters of the harmonic potential as $(a^{(1)}_{t_i}, a^{(2)}_{t_i}) = (a_0, a_0)$ and $(a^{(1)}_{t_f}, a^{(2)}_{t_f})  = (2a_0, 2a_0)$, respectively. 
The corresponding symmetric process is realized by choosing 
$a_{t_i} = a_0$ and $a_{t_f} = 2a_0$. 
Then, from Eqs.\ (\ref{eqn:irr_w_sym}) and (\ref{eqn:mean_e_sym_large}), 
the mean work and the change of the mean energy in the large $\tau_{op}$ limit are given by 
\begin{eqnarray}
\lim_{\tau_{op} \rightarrow \infty} \frac{W_{sym}}{W^{(1)}} 
&=& 
1+ 
\frac{ \left( \sqrt{2} - 1 \right)^2}{\ln 2} 
\frac{\tau^*}{\tau_{op}} 
\, , \label{eqn:app_mw_sym_tinf} \\
\lim_{\tau_{op} \rightarrow \infty} \frac{ \Delta E_{sym}}{E(t_i)} 
&=& 
\frac{\left( \sqrt{2} - 1 \right)}{2}  \frac{\tau^*}{\tau_{op}} \, , \label{eqn:e_largetau}
\end{eqnarray}
respectively. Here we introduced $\tau^* = \nu /a_0$.

\section{Application (II) : Asymmetric process} \label{sec:asym_prot}

\begin{figure}[h]
\includegraphics[scale=0.4]{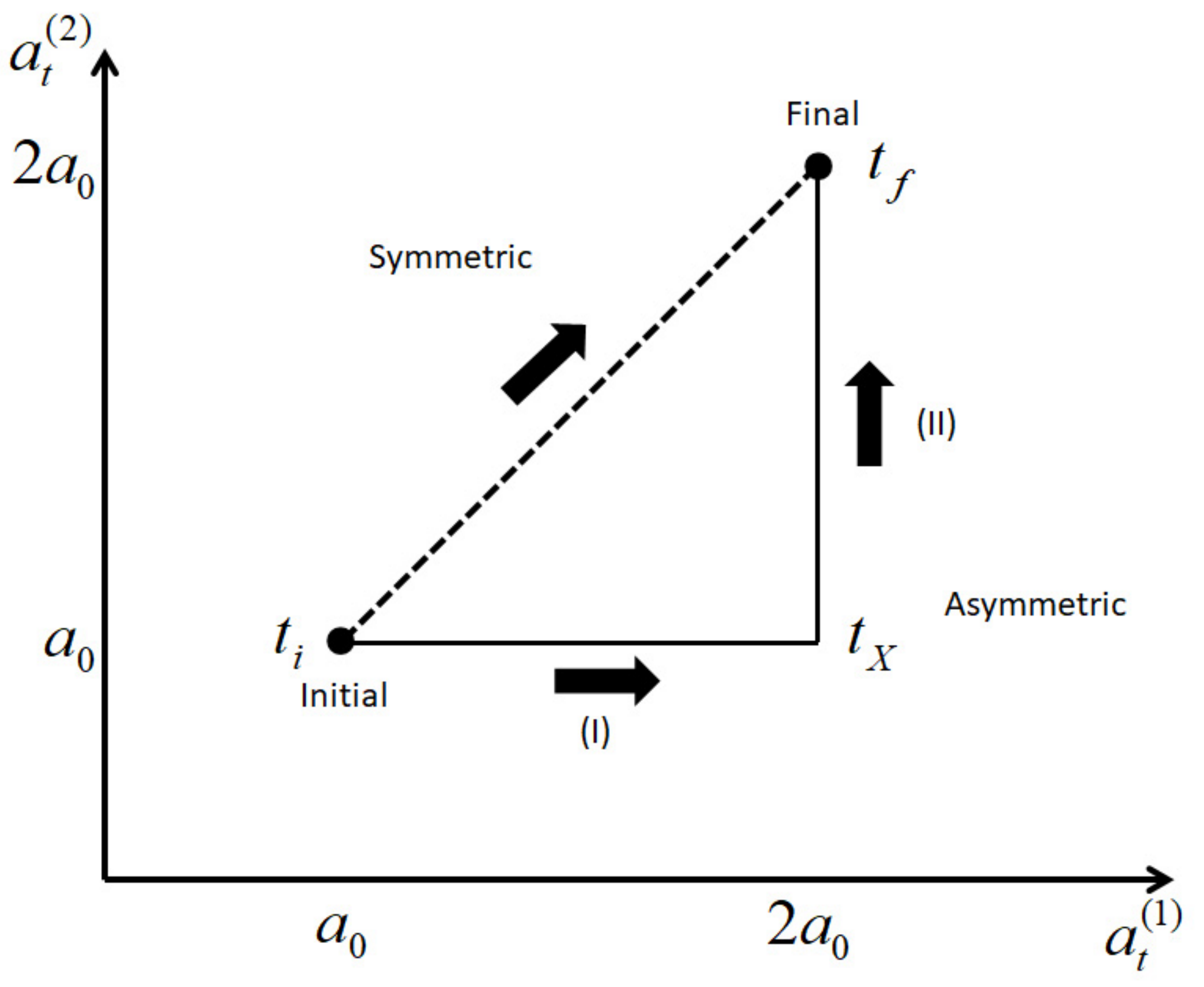}
\caption{The symmetric and asymmetric compression processes of the harmonic potential. 
The former is denoted by the dotted line and the latter by the solid line, respectively. 
The rotational symmetry of the harmonic potential is hold in the symmetry process where $a^{(1)}_t = a^{(2)}_t$. 
The asymmetric process is represented by two paths. 
The control parameter $a^{(2)}_t$ is fixed by $a_0$ on the path (I). 
At the switching time $t_X$, the path (I) is switched to the path (II) where $a^{(1)}_t$ is fixed by $2a_0$.}
\label{fig:path}
\end{figure}

In this section, we consider the asymmetric deformation of the harmonic potential where $a^{(1)}_t \neq a^{(2)}_t$ for $t_i < t < t_f$.
The initial and final values of the control parameters of the harmonic potential are fixed by $(a^{(1)}_{t_i},a^{(2)}_{t_i}) = (a_0,a_0)$ and $(a^{(1)}_{t_f},a^{(2)}_{t_f}) = (2a_0,2a_0)$, respectively. 
We first increase $a^{(1)}_t$ fixing $a^{(2)}_t = a_0$ along the path (I) as shown in Fig.\ \ref{fig:path}.
After the switching time $t_X$, the control parameters are changed along the path (II) where $a^{(2)}_t$ is increased fixing $a^{(1)}_t=2a_0$.
That is, the corresponding time evolutions of the control parameters are represented by  
\begin{eqnarray}
(a^{(1)}_{t_f},a^{(2)}_{t_f}) =
\left\{
\begin{array}{cl}
(f_t , a_0) & : {\rm path\, (I)}\, \, \, \, \, t_i \le t \le t_X \\
(2a_0 , g_t) &: {\rm path\, (II)}\, \, \, \, \, t_X < t \le t_f 
\end{array}
\right. \, ,
\end{eqnarray}
where $f_t$ and $g_t$ are monotonically increasing functions satisfying 
\begin{eqnarray}
f_{t_i} = g_{t_X} =a_0 \, , \,\,\,\,\,\,\,\,\, f_{t_X} = g_{t_f} =2a_0 \, .
\end{eqnarray}
The switching time $t_X$ and the functions $f_t$ and $g_t$ are determined by minimizing the mean work.

%Differently from the symmetric process, the degeneracy, $\ud^{(n)}({\bf a}_t)$, is a function of time.
To apply the perturbation theory to this case, it is convenient to introduce 
the states which are parameterized by the eigenvalues of each spatial component,  
\begin{eqnarray}
\left\{
\begin{split}
\langle {\bf x} | \underline{``{\bf l}", {\bf a}_t} \rangle &=
\langle {\bf x} | \underline{``l_1,l_2", {\bf a}_t} \rangle = \underline{\rho}_{l_1} (x_1, a^{(1)}_t) \underline{\rho}_{l_2} (x_2, a^{(2)}_t)  \\
\langle {\bf x} | \overline{``{\bf l}", {\bf a}_t} \rangle &=
\langle {\bf x} | \overline{``l_1,l_2", {\bf a}_t} \rangle = \overline{\rho}_{l_1} (x_1, a^{(1)}_t) \overline{\rho}_{l_2} (x_2, a^{(2)}_t) 
\end{split}
\right. 
\label{eqn:oldpara} \, ,
\end{eqnarray}
where we have used the functions defined in Eqs.\ (\ref{eqn:ho_foot}) and (\ref{eqn:ho_head}).  
The corresponding eigenvalues of the Fokker-Planck operator are characterized by the pair of the two integers $(l_1,l_2)$, 
\begin{eqnarray}
\lambda_{(l_1,l_2)} (t) = \frac{a^{(1)}_t}{\nu} l_1 + \frac{a^{(2)}_t}{\nu} l_2 \, .
\end{eqnarray}
These states satisfy the orthonormal condition,
\begin{eqnarray}
\langle \overline{``{\bf m}",{\bf a}_t} | \underline{``{\bf n}",{\bf a}_t} \rangle 
=
\delta_{n_1,m_1} \delta_{n_2,m_2}
=
\delta_{{\bf n}, {\bf m}} \, .
\end{eqnarray}
We call these the fictitious states.
The sums associated with $n$ and $\alpha_n$, 
are shown to be equivalent to the sums of $l_1$ and $l_2$, 
\begin{eqnarray}
\sum_{n \ge 0} \sum_{\greekbf{\alpha}_n =0}^{{\bf d}^{(n)}({\bf a}_t)} = \sum_{l_1\ge 0} \sum_{l_2 \ge 0} \, ,
\end{eqnarray}
and hence the condition (\ref{eqn:time-ind-sum}) is satisfied.
It is however noted that the degeneracy of the fictitious state is not yet identified.

Because of the same reason discussed below Eq.\ (\ref{eqn:worksum}), 
the exact mean work is given by the sum of $W^{(1)}$ and $W^{(2)}_{asym}$.
As discussed, $W^{(1)}$ is independent of the evolution of the control parameters and given by Eq.\ (\ref{eqn:w(1)}),
$W^{(1)} = \frac{1}{\beta} \ln 2$.
The second order term $W^{(2)}_{asym} $ is given by the sum of the contributions from the path (I) and the path (II), 
\begin{eqnarray}
W^{(2)}_{asym} 
= 
W^{(2)}_{(I)} + W^{(2)}_{(II)} \, ,
\end{eqnarray}
where   
\begin{eqnarray}
W^{(2)}_{(I)} 
&=& 
- \sum_{n \ge 1}\int^{t_X}_{t_i} \ud t \int^{t}_{t_i} \ud s \, 
\sum_{\alpha_n,\beta_n=0}^{d^{(n)}({\bf a}_t)}
\sum_{\gamma_n=0}^{d^{(n)}({\bf a}_s)}
 e^{-\int^t_{s} \ud s^\prime\, \lambda_2 (s^\prime) } 
\langle \overline{0,0, {\bf a}_i} |(\partial_t \widehat{V})| \underline{n,\alpha_n,{\bf a}_t} \rangle 
\nonumber \\
&& \times 
\frac{
[{\bf M}^{(n)}(t) ]_{\alpha_n,\beta_n} 
{[{\bf M}^{(n)}(s) ]^{-1}}_{\beta_n,\gamma_n} 
\left[\dot{\bf L}^{(n,0)}(s) \right]_{\gamma_n,0}
[{\bf M}^{(0)}(s)]_{0,0}
}{\lambda_n(s)}  \label{eqn:w2i}\, ,\\
W^{(2)}_{(II)}
&=&
- \sum_{n\ge 1}\int^{t_f}_{t_X} \ud t \int^{t}_{t_i} \ud s \, 
\sum_{\alpha_n,\beta_n=0}^{d^{(n)}({\bf a}_t)}
\sum_{\gamma_n=0}^{d^{(n)}({\bf a}_s)}
 e^{-\int^t_{s} \ud s^\prime\, \lambda_2 (s^\prime) } 
\langle \overline{0,0, {\bf a}_i} |(\partial_t \widehat{V})| \underline{n,\alpha_n,{\bf a}_t} \rangle 
\nonumber \\
&& \times 
\frac{
[{\bf M}^{(n)}(t) ]_{\alpha_n,\beta_n} 
{[{\bf M}^{(n)}(s) ]^{-1}}_{\beta_n,\gamma_n} 
\left[\dot{\bf L}^{(n,0)}(s) \right]_{\gamma_n,0}
[{\bf M}^{(0)}(s)]_{0,0}
}{\lambda_n(s)} \label{eqn:w2ii} \, .
\end{eqnarray}
We see that the integration of $t$ in $W^{(2)}_{(I)}$ is on the path (I) while that in $W^{(2)}_{(II)}$ is on the path (II).

\subsection{Calculation of $W^{(2)}_{(I)}$}

\begin{figure}[h]
\begin{minipage}{0.4\textwidth}
\begin{center}
\hspace{-2cm}
\includegraphics[width=80mm]{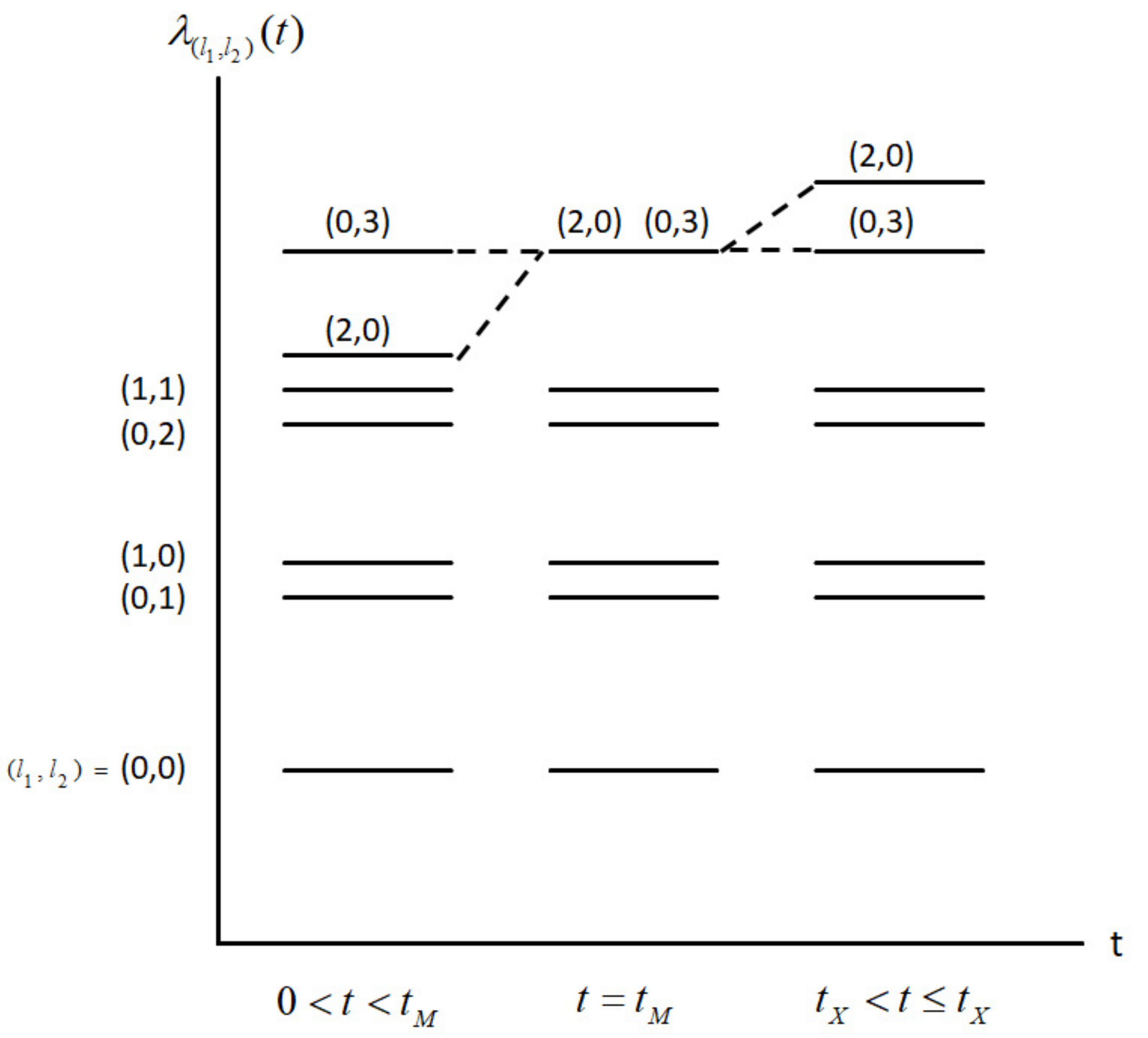}
\end{center}
\end{minipage}
\begin{minipage}{0.4\textwidth}
\begin{center}
\makeatletter
\def\@caption{Classification of degeneracy of eigenvalue and eigenfunction for $t_i < t \le t_X$}
\makeatletter
\begin{tabular}{|c|c|c|c|}
\hline
$|\underline{n,\alpha_n,{\bf a}_t} \rangle$ & ~$\lambda_n(t)$~ & $|\underline{``l_1,l_2"{\bf a}_t} \rangle$ & \\
\hline \hline
$|\underline{0,0,{\bf a}_t} \rangle$ & $0$ &   $| \underline{``0,0,"{\bf a}_t} \rangle $    &\\
\hline
$|\underline{1,0,{\bf a}_t} \rangle$ & $\frac{a_0}{\nu}$ &   $| \underline{``0,1," {\bf a}_t} \rangle $    &\\
\hline
$|\underline{2,0,{\bf a}_t} \rangle$ & $\frac{a_t}{\nu}$ &   $| \underline{``1,0," {\bf a}_t} \rangle $    &\\
\hline
$|\underline{3,0,{\bf a}_t} \rangle$ & $\frac{2a_0}{\nu}$ &   $| \underline{``0,2," {\bf a}_t} \rangle $   &\\
\hline
$|\underline{4,0,{\bf a}_t} \rangle$ & $\frac{a_t}{\nu}+ \frac{a_0}{\nu}$ &   $| \underline{``1,1," {\bf a}_t} \rangle $   &\\
\hline
$|\underline{5,0,{\bf a}_t} \rangle$ & $\frac{2a_t}{\nu} $ &   $| \underline{``2,0," {\bf a}_t} \rangle $   & ~$t_i < t < t_M$~ \\
\hline
$|\underline{5,0,{\bf a}_t} \rangle$ &\multirow{2}{*}{$\frac{3a_0}{\nu}$} &   $| \underline{``2,0," {\bf a}_t} \rangle $  &  \multirow{2}{*}{$t=t_M$} \\
$|\underline{5,1,{\bf a}_t} \rangle$ &  &   $| \underline{``0,3," {\bf a}_t} \rangle $  & \\
\hline
$|\underline{5,0,{\bf a}_t} \rangle$ & $\frac{3a_0}{\nu} $ &   $| \underline{``0,3," {\bf a}_t} \rangle $  & $ t_M < t \le t_X $ \\
\hline
$|\underline{6,0,{\bf a}_t} \rangle$ & $\frac{3a_0}{\nu}$ & $| \underline{``0,3,"{\bf a}_t} \rangle $  &  $t_i < t < t_M$ \\
\hline
$|\underline{6,0,{\bf a}_t} \rangle$ & \multirow{2}{*}{$\frac{4a_0}{\nu}$}  & $| \underline{``2,1,"{\bf a}_t} \rangle $   & \multirow{2}{*}{$t=t_M$}  \\
$|\underline{6,1,{\bf a}_t} \rangle$ &  &  $| \underline{``0,4,"{\bf a}_t} \rangle $    &\\
\hline
$|\underline{6,0,{\bf a}_t} \rangle$ & $\frac{2a_t}{\nu} $ &   $| \underline{``2,0,"{\bf a}_t} \rangle $   & $t_M < t \le t_X$ \\
\hline
\end{tabular}
\label{table1}
\end{center}
\end{minipage}
\caption{The distribution of the eigenvalues of the fictitious states (\ref{eqn:oldpara}) for $t_i < t \le t_X$.
The left panel shows the schematic figure of the change of the eigenvalues. 
The right panel represents the correspondence between the foot state $| \underline{n,\alpha_n, {\bf a}_t} \rangle$ 
and the fictitious state $|\underline{``l_1,l_2",{\bf a}_t} \rangle$. 
}
\label{fig:degeneracy}
\end{figure}

To calculate these equations, we have to identify a fictitious state $|\underline{``l_1,l_2",{\bf a}_t} \rangle$ with a foot state $| \underline{n,\alpha_n, {\bf a}_t} \rangle$. 
We first consider the contribution on the path (I). 
On this path, the correspondence between the two states are summarized in Fig.\ \ref{fig:degeneracy}. 
We find that, for example, the sixth foot state $| 5,0, {\bf a}_t \rangle$ corresponds to $| ``2,0" {\bf a}_t \rangle$ at $t_i < t <t_M$ 
and to $| ``3,0" {\bf a}_t \rangle$ at $t_M < t < t_f$. 
Here the time $t_M$ is defined by 
\begin{eqnarray}
a_{t_M} = \frac{3}{2} a_0 \, .
\end{eqnarray}
This foot state is degenerate instantaneously at $t=t_M$ and hence
\begin{eqnarray}
\ud^{(5)} ({\bf a}_t) 
= \left\{
\begin{array}{cc}
0 & t \neq t_M \, ,\\
1 & t=t_M  \, .
\end{array}
\right. 
\end{eqnarray}
In the calculation of the mean work, the integrand is not singular and thus the contribution at the moment $t=t_M$ is negligible.

From the correspondence in  Fig.\ \ref{fig:degeneracy}, 
we observe, for example, that the fictitious state $(l_1,l_2) = (2,0)$ corresponds to the sixth foot state $(n,\alpha_n) = (5,0)$ for $t_i < t < t_M$, 
but it is changed to the seventh foot state $(n,\alpha_n) = (6,0)$ for $t_M < t \le t_X$. 
Because of this change of the eigenvalues, $W^{(2)}_{(I)}$ is given by the sum of three contributions, 
\begin{eqnarray}
W^{(2)}_{(I)}
&=&
W^A_{(I)} + W^B_{(I)} + W^C_{(I)}  \nonumber \\
&=&
\int^{t_X}_{t_i} \ud t  \int^{t}_{t_i} \ud s \,   
 \frac{\ud f_s}{\ud s}
\left[ e^{-\int^t_{s} \ud s^\prime\, \frac{2f_{s^\prime}}{\nu} } 
 \frac{1}{2 \beta  (f_s)^2} 
\right] \frac{\ud f_t}{\ud t}
\, , \label{eqn:w2part1}
\end{eqnarray}
where
\begin{eqnarray}
W^A_{(I)} 
&=&
- \int^{t_M}_{t_i} \ud t \int^{t}_{t_i} \ud s \, 
%\sum_{\beta_5=0}^{d^{(5)} ({\bf a}_t)} \sum_{\gamma_5=0}^{d^{(5)} ({\bf a}_\tau)} 
e^{-\int^t_{s} \ud s^\prime\, \lambda_{(2,0)} (s^\prime) } \nonumber \\
&& \times 
\frac{[{\bf M}^{(5)} (t)]_{0, 0}  [{\bf M}^{(5)} (s)]^{-1}_{0, 0} [\dot{\bf L}^{(5,0)} (s)]_{0,0} [{\bf M}^{(0)} (s)]_{0,0}}{\lambda_{(2,0)} (s)}
\langle \overline{ 0,0,{\bf a}_{t_i} } | (\partial_t \widehat{V})
| \underline{ 5, 0, {\bf a}_t } \rangle  \nonumber \\
&=& 
 \int^{t_M}_{t_i} \ud t \int^{t}_{t_i} \ud s \, 
 \frac{\ud f_s}{\ud s} \left[ e^{-\int^t_{s} \ud s^\prime\, \frac{2 f_{s^\prime}}{\nu} }
 \frac{1 }{2\beta (f_s)^2} 
\right] \frac{\ud f_t}{\ud t}
\, , \\
W^B_{(I)} 
&=& -  \int^{t_X}_{t_M} \ud t  \int^{t_M}_{t_i} \ud s \,   
 e^{-\int^t_{s}\ud s^\prime\, \lambda_{(2,0)} (s^\prime) }\nonumber \\
&& \times 
\frac{[{\bf M}^{(6)} (t)]_{0, 0} [{\bf M}^{(5)} (s)]^{-1}_{0, 0} [\dot{\bf L}^{(5,0)} (s)]_{0,0} [{\bf M}^{(0)} (s)]_{0,0}}{\lambda_{(2,0)} (s)}
\langle \overline{ 0,0,{\bf a}_{t_i} } | (\partial_t \widehat{V})
| \underline{ 6, 0, {\bf a}_t } \rangle  \nonumber \\
&=&
\int^{t_X}_{t_M} \ud t  \int^{t_M}_{t_i} \ud s \,   
 \frac{\ud f_s}{\ud s} 
\left[ e^{-\int^t_{s} \ud s^\prime\, \frac{2f_{s^\prime}}{\nu} } 
 \frac{1}{2 \beta  (f_s)^2} 
\right] \frac{\ud f_t}{\ud t}
\, ,  \label{eqn:wb1}\\
W^C_{(I)} 
&=& - \int^{t_X}_{t_M} \ud t \int^{t}_{t_M} \ud s  \,   
 e^{-\int^t_{s} \ud s^\prime\, \lambda_{(2,0)} (s^\prime) }\nonumber \\
&& \times 
\frac{[{\bf M}^{(6)} (t)]_{0, 0}  [{\bf M}^{(6)} (s)]^{-1}_{0, 0} [\dot{\bf L}^{(6,0)} (s)]_{0,0} [{\bf M}^{(0)} (s)]_{0,0}}{\lambda_{(2,0)} (s)}
\langle \overline{ 0,0,{\bf a}_{t_i} } | (\partial_t \widehat{V})
| \underline{ 6, 0, {\bf a}_t } \rangle  \nonumber \\
&=&
 \int^{t_X}_{t_M} \ud t  \int^{t}_{t_M} \ud s \,   
 \frac{\ud f_s}{\ud s}
\left[ e^{-\int^t_{s} \ud s^\prime\, \frac{2f_{s^\prime}}{\nu} } 
 \frac{1}{2 \beta  (f_s)^2} 
\right]  \frac{\ud f_t}{\ud t}
\, .
\end{eqnarray}
In the calculations of $W^A_{(I)}$ and $W^C_{(I)}$, 
there is no effect of the change of the eigenvalues. 
We, however, require attention to the calculation of $W^{B}_{(I)}$.
The fictitious state $(l_1,l_2) = (2,0)$ in $W^{(B)}_{(I)}$ corresponds to the foot state $(n,\alpha_n) = (5,0)$ for the integral of $s$ 
but the same state represents $(n,\alpha_n) = (6,0)$ for the integral of $t$. 
This is the reason why we observe $[{\bf M}^{(6)} (t)]_{0, 0}$ and $ [{\bf M}^{(5)} (s)]^{-1}_{0, 0}$ in Eq.\ (\ref{eqn:wb1}) at the same time,
\begin{eqnarray}
\lefteqn{\sum_{n\ge 1} \int^{t_X}_{t_M} \ud t\, \int^{t_M}_{t_i} \ud s \, [{\bf M}^{(n)}(t)]_{0,0}[{\bf M}^{(n)}(s)]^{-1}_{0,0}
\langle \overline{ 0,0,{\bf a}_s } | 
(\partial_t \widehat{V})
| \underline{ n, \alpha_n, {\bf a}_t } \rangle } && \nonumber \\
&&= 
 \int^{t_X}_{t_M} \ud t\, \int^{t_M}_{t_i} \ud s \, [{\bf M}^{(6)}(t)]_{0,0}[{\bf M}^{(5)}(s)]^{-1}_{0,0}
\langle \overline{ 0,0,{\bf a}_s } | 
(\partial_t \widehat{V})
| \underline{ 6, 0, {\bf a}_t } \rangle \, .
\end{eqnarray}

\subsection{Calculation of $W^{(2)}_{(II)}$}

\begin{figure}[h]
\begin{center}
\begin{tabular}{|c|c|c|}
\hline
$| \underline{n, \alpha_n, {\bf a}_t} \rangle $  & $\lambda_n(t)$ & $| \underline{``l_1,l_2,"{\bf a}_t} \rangle $    \\
\hline \hline
$| \underline{0, 0, {\bf a}_t} \rangle $ & $0$ &   $| \underline{``0,0,"{\bf a}_t} \rangle $    \\
\hline
$| \underline{1,0, {\bf a}_t} \rangle $ & $\frac{a_t}{\nu}$ &   $| \underline{``0,1,"{\bf a}_t} \rangle $    \\
\hline
$| \underline{2,0, {\bf a}_t} \rangle $ & $\frac{2 a_0}{\nu}$ &   $| \underline{``1,0,"{\bf a}_t} \rangle $     \\
\hline
$| \underline{3,0, {\bf a}_t} \rangle $ & $\frac{2a_t}{\nu}$ &  $| \underline{``0,2,"{\bf a}_t} \rangle $     \\
\hline
$| \underline{4, 0, {\bf a}_t} \rangle $ & $\frac{2a_0}{\nu}+ \frac{a_t}{\nu}$ &  $| \underline{``1,1,"{\bf a}_t} \rangle $     \\
\hline
$| \underline{5,0, {\bf a}_t} \rangle $ & $\frac{4a_0}{\nu}$ & $| \underline{``2,0,"{\bf a}_t} \rangle $   \\
\hline
$| \underline{6, 0, {\bf a}_t} \rangle $ & $\frac{2a_0}{\nu} + \frac{2a_t}{\nu}$ & $| \underline{``1,2,"{\bf a}_t} \rangle $   \\
\hline
\end{tabular}
\caption{The distribution of the eigenvalues of the fictitious state (\ref{eqn:oldpara}) for $t_X < t < t_f$.}
\label{table2}
\end{center}
\end{figure}

We consider the contribution from the path $(II)$ in this subsection. 
The distribution of the eigenvalues on the path (II) is summarized in Fig.\ \ref{table2}, which is much simpler than 
that of the path (I).
In this case, Eq.\ (\ref{eqn:w2ii}) is calculated as
\begin{eqnarray}
W^{(2)}_{II} 
&=& - \int^{t_f}_{t_X} \ud t \int^{t}_{t_i}   \ud s \, 
 e^{-\int^t_{s} \ud s^\prime\, \lambda_3 (s^\prime) } 
\frac{
[{\bf M}^{(3)}(t) ]_{0,0} 
{[{\bf M}^{(3)}(s) ]^{-1}}_{0,0} 
\left[\dot{\bf L}^{(3,0)}(s) \right]_{0,0}
}{\lambda_n(s)} 
\langle \overline{0,0, {\bf a}_s} |(\partial_t \widehat{V})| \underline{3,0,{\bf a}_t} \rangle \nonumber \\
&=& - \int^{t_f}_{t_X} \ud t \int^{t}_{t_X}   \ud s \, 
 e^{-\int^t_{s} \ud s^\prime\, \lambda_3 (s^\prime) } 
\frac{
[{\bf M}^{(3)}(t) ]_{0,0} 
{[{\bf M}^{(3)}(s) ]^{-1}}_{0,0} 
\left[\dot{\bf L}^{(3,0)}(s) \right]_{0,0}
}{\lambda_n(s)} 
\langle \overline{0,0, {\bf a}_s} |(\partial_t \widehat{V})| \underline{3,0,{\bf a}_t} \rangle 
 \nonumber \\
&=& 
  \int^{t_f}_{t_X} \ud t \int^{t}_{t_X}  \ud s \, 
\frac{\ud g_s}{\ud s}
\left[ e^{-\int^t_{s} \ud s^\prime\, \frac{2g_{s^\prime}}{\nu} } 
\frac{1}{2\beta (g_s)^2 } \right] \frac{\ud g_t}{\ud t} \, .
\label{eqn:w2part2}
\end{eqnarray}
In the second equality, the lower limit of the integral of $s$ is changed from $t_i$ to $t_X$ because 
\begin{eqnarray}
\dot{\bf L}^{(n,0)} (s) 
&=&
- \frac{\sqrt{2}}{\nu} 
\left\{
\begin{array}{cc}
 \frac{\ud f_s}{\ud s} 
\sqrt{\frac{N^{(1)}_0 (s)}{N^{(1)}_2 (s)}} 
\delta_{n,5}\delta_{\alpha_n,0} \,\,
& t_i < s < t_M \, ,\\
 \frac{\ud f_s}{\ud s} 
\sqrt{\frac{N^{(1)}_0 (s)}{N^{(1)}_2 (s)}} 
\delta_{n,6}\delta_{\alpha_n,0} \,\,
& t_M < s < t_X \, ,
\\
\frac{\ud g_s}{\ud s} 
\sqrt{\frac{N^{(2)}_0 (s)}{N^{(2)}_2 (s)}} 
\delta_{n,3}\delta_{\alpha_n,0} \,\,
& t_X < s < t_f  \, .
\end{array}
\right. 
\end{eqnarray}

\subsection{Optimization in Asymmetric Process} \label{sec:opt_asym}

The second order contribution of the mean work in the asymmetric process is given by the sum of Eqs.\ (\ref{eqn:w2part1}) and (\ref{eqn:w2part2}), 
\begin{eqnarray}
W_{asym}^{(2)}
&=& 
 \int^{t_X}_{t_i} \ud t  \int^{t}_{t_i}\ud s \,   
\frac{\ud f_s}{\ud s}
\left[ e^{-\int^t_{s} \ud s^\prime\, \frac{2f_{s^\prime}}{\nu} } 
 \frac{1}{2 \beta  (f_s)^2} 
\right] \frac{\ud f_t}{\ud t} \nonumber \\
&&+ 
 \int^{t_f}_{t_X} \ud t \int^{t}_{t_X}  \ud s \, 
\frac{\ud g_s}{\ud s} \left[ e^{-\int^t_{s} \ud s^\prime\, \frac{2g_{s^\prime}}{\nu} } 
\frac{1}{2\beta (g_s)^2 } \right] \frac{\ud g_t}{\ud t} \, . 
\label{eqn:wirr_asym}
\end{eqnarray}
As was done in Sec.\ \ref{sec:opt}, 
the mean work in the the large $\tau_{op}$ limit is minimized to find the optimized process.
In this limit, $W^{(2)}_{asym}$ is given by 
\begin{eqnarray}
\lim_{\tau_{op} \rightarrow \infty} 
W_{asym}^{(2)} 
= 
\frac{\nu}{4 \beta a_0 \tau_{op}}
\int^{\tau_X}_{0} \ud \tau \, 
\frac{1}{ (\bar{f}_{\tau})^3} \left( \frac{\ud \bar{f}_{\tau}}{\ud \tau} \right)^2
 + 
\frac{\nu}{4 \beta a_0 \tau_{op}}
 \int^{1}_{\tau_X} \ud \tau \, 
\frac{1}{ (\bar{g}_{\tau})^3}
\left( \frac{\ud \bar{g}_{\tau}}{\ud \tau} \right)^2
 \, , \label{eqn:wir_1+2_app}
\end{eqnarray}
where adimensional quantities are introduced by 
\begin{eqnarray}
\tau_X &=& (t_X-t_i)/\tau_{op} \, , \\
\bar{f}_\tau &=& \frac{f_{t_i + \tau \tau_{op}}}{a_0} \, , \\
\bar{g}_\tau &=& \frac{g_{t_i + \tau \tau_{op}}}{a_0} \, .
\end{eqnarray}
The optimized parameters $\bar{f}_{\tau}$ and $\bar{g}_{\tau}$ 
are obtained by minimizing $W^{(2)}_{asym}$.  
The variations of $f_t$ and $g_t$ are defined by 
\begin{eqnarray}
\bar{f}_\tau &\longrightarrow& \bar{f}_\tau + \delta \bar{f}_\tau \, ,\\
\bar{g}_\tau &\longrightarrow& \bar{g}_\tau + \delta \bar{g}_\tau \, ,
\end{eqnarray}
which satisfy 
$\delta \bar{f}_0 = \delta \bar{f}_{\tau_X} = \delta \bar{g}_{\tau_X} = \delta \bar{g}_1 = 0$. 
The optimized parameters are given by the solutions of the following differential equations, 
\begin{eqnarray}
\frac{1}{\bar{f}^3_{\tau}} \frac{\ud^2 \bar{f}_\tau }{\ud \tau^2}- \frac{3}{2} \frac{1}{\bar{f}^4_\tau} 
\left( \frac{\ud \bar{f}_\tau}{\ud \tau}\right)^2 &=& 0 \, ,\\
\frac{1}{\bar{g}^3_{\tau}} \frac{\ud^2 \bar{g}_\tau }{\ud \tau^2}- \frac{3}{2} \frac{1}{\bar{g}^4_\tau} 
\left( \frac{\ud \bar{g}_\tau}{\ud \tau}\right)^2 &=& 0 \, .
\end{eqnarray}
These equations are solved under the following conditions:
\begin{eqnarray}
\bar{f}_0 = \bar{g}_X = 1\, , \,\,\,\,\,\,\,\bar{f}_X = \bar{g}_1 = 2 \, .
\end{eqnarray}
The solutions are given by 
\begin{eqnarray}
\begin{split}
\bar{f}_\tau = \frac{2\tau^2_X}{(\sqrt{2}\tau_X -(\sqrt{2}-1)\tau)^2} &\longrightarrow 
f_t = \frac{2(t_X-t_i)^2}{\{ (t-t_i) + (t_X - t)\sqrt{2} \}^2}a_0 \, ,   \\
\bar{g}_\tau = \frac{2(1-\tau_X)^2}{(\tau_X - \sqrt{2} + (\sqrt{2}-1) \tau)^2} &\longrightarrow
g_t = \frac{2(t_f-t_X)^2}{\{ (t-t_X) + (t_f - t)\sqrt{2} \}^2}a_0 \, . 
\end{split}
\label{eqn:opt_profg}
\end{eqnarray}

The optimized switching time $t_X$ is determined 
by minimizing $W^{(2)}_{asym}$ which is obtained by substituting Eq.\ (\ref{eqn:opt_profg}) into Eq.\ (\ref{eqn:wir_1+2_app}). 
Solving $\partial W^{(2)}_{asym}/\partial \tau_X = 0$, 
the optimized $t_X$ is given by 
\begin{eqnarray}
\tau_X = \frac{1}{2} \longrightarrow t_X = \frac{t_f + t_i}{2}\, . \label{eqn:tx}
\end{eqnarray}

Using these results, the optimized mean work in the large $\tau_{op}$ limit is eventually given by 
\begin{eqnarray}
\lim_{\tau_{op}\rightarrow \infty} 
\frac{W_{asym}}{W^{(1)}} 
= 
1+
 2 \frac{(\sqrt{2}-1)^2}{\ln 2} 
\frac{\tau^*}{\tau_{op}}
 \, , \label{eqn:app_mw_asym_tinf}
\end{eqnarray}
where $\tau^* = \nu/a_0$ which is already defined below Eq.\ (\ref{eqn:e_largetau}).

In a similar fashion, the change of the mean energy in the asymmetric process is calculated by 
applying the above results to the definition (\ref{eqn:def_ener}) as was done in Sec.\ \ref{sec:ener},
\begin{eqnarray}
\Delta E^{asym}
=
\frac{f_{t_X}}{2\beta} \int^{t_X}_{t_i} d\tau \,   
\left[ e^{-\int^{t_X}_{\tau}ds\, \frac{2f_s}{\nu} } 
 \frac{1}{(f_\tau)^2} 
\right] \frac{\ud f_t}{\ud t}
+ 
 \frac{g_{t_f}}{2\beta}  \int^{t_f}_{t_X}  d\tau \, 
\left[ e^{-\int^{t_f}_{\tau}ds\, \frac{2g_s}{\nu} } 
\frac{1}{ (g_\tau)^2 } \right] \frac{\ud g_t}{\ud t} \, . 
\label{eqn:deltae_asym}
\end{eqnarray}
Using Eqs.\ (\ref{eqn:opt_profg}) and (\ref{eqn:tx}) in the large $\tau_{op}$ limit, 
the optimized change of the mean energy is calculated by 
\begin{eqnarray}
\lim_{\tau_{op} \rightarrow \infty}\frac{\Delta E^{asym}}{E(t_i)}
= 
(\sqrt{2}-1)  \frac{\tau^*}{\tau_{op}} \, .
\label{eqn:e_largetau_asym}
\end{eqnarray}

\section{numerical results} \label{sec:numerical}

We compare the mean works and the changes of the mean energy in the symmetric and asymmetric compression processes 
which are denoted by the dotted and solid lines in Fig.\ \ref{fig:path}, respectively. 
The initial and final values of the control parameters are fixed by
$(a^{(1)}_{t_i},a^{(2)}_{t_i}) = (a_0,a_0)$ and $(a^{(1)}_{t_f},a^{(2)}_{t_f}) = (2a_0,2a_0)$ in both processes.
We use the optimized control parameters which are given by Eq.\ (\ref{eqn:app_protocol}) in the symmetric process, 
and by Eqs.\ (\ref{eqn:opt_profg}) and (\ref{eqn:tx}) in the asymmetric process.

\subsection{Mean Work}

\begin{figure}[t]
\begin{center}
%\vspace*{-3cm}
\includegraphics[scale=0.3]{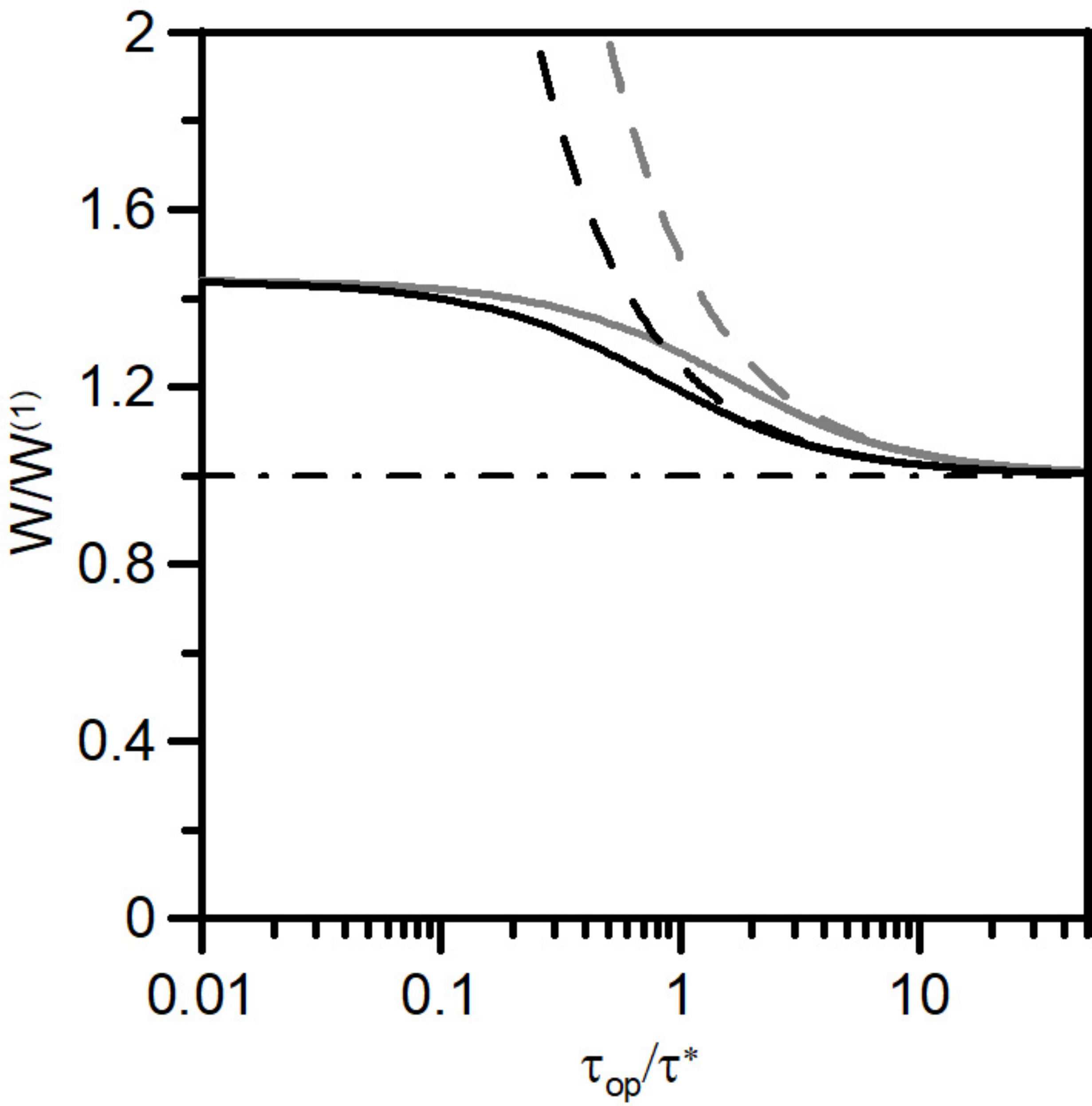}
\end{center}
\caption{The mean works are plotted as functions of the operation time $\tau_{op}=t_f - t_i$.
The black solid and dotted lines are results in the symmetric process and 
represent the exact mean work and the mean work in the large $\tau_{op}$ limit, respectively. 
The optimized control parameters are given by Eq.\ (\ref{eqn:app_protocol}).
The gray solid and dotted lines are results in the asymmetric process and denote the exact mean work  
and the mean work in the large $\tau_{op}$ limit, respectively.
The optimized parameters are given by Eq.\ (\ref{eqn:opt_profg}).
The dotted-dashed line represents $W^{(1)}$ which is given by Eq.\ (\ref{eqn:w(1)}).
We define $\tau^*=\nu/a_0$.
 }
\label{fig:mw}
\end{figure}

In Fig.\ \ref{fig:mw}, the mean works are plotted as functions of the operation time $\tau_{op}= t_f - t_i$.
The black and gray solid lines represent the exact mean works in the symmetric 
and asymmetric processes, respectively. 
The former is calculated from Eqs.\  (\ref{eqn:exa_irw_2d}) and (\ref{eqn:w(1)}), 
and the latter from Eqs.\ (\ref{eqn:w(1)}) and (\ref{eqn:wirr_asym}).
The black and gray dotted lines show the mean works in the large $\tau_{op}$ limit which are 
given by Eqs.\ (\ref{eqn:app_mw_sym_tinf}) and (\ref{eqn:app_mw_asym_tinf}), respectively.
We find that the asymptotic behaviors of the solid lines are well reproduced by the corresponding dotted lines. 
In the large $\tau_{op}$ limit, these processes correspond to the quasi-static process. 
Then the black and gray lines converge to $W^{(1)}$, 
which corresponds to the mean work in the reversible process.

In the instantaneous jump limit $\tau_{op} \rightarrow 0$, the black and gray solid lines converge to the same value, 
\begin{eqnarray}
\lim_{\tau_{op}\rightarrow 0} W^{sym} 
=
\lim_{\tau_{op}\rightarrow 0} W^{asym}  
= \frac{1}{\beta} 
\label{eqn:w_instantaneous}
\, . 
\end{eqnarray}
This behavior can be understood from the first law (\ref{eqn:se_1law}).
In this limit, there is no enough time for a Brownian particle to interact with the thermal bath 
and thus Eq.\  (\ref{eqn:se_1law}) leads to $\ud W_t = E(t+\ud t ) - E(t)$  because $\ud Q_t =0$.
Moreover, in this limit, the final states are approximately given by the initial state, $\rho({\bf x},t_f) = \rho({\bf x},t_i)$.
Therefore the changes of the mean energy are given by $1/\beta$ both in the symmetric and asymmetric processes, 
\begin{eqnarray}
E(t_f) - E(t_i) 
&=&  \int \ud^2 {\bf x} \, \rho({\bf x},t_f) V({\bf x},a_f) - \int \ud^2 {\bf x} \, \rho({\bf x},t_i) V({\bf x},a_i)  \nonumber \\
&=&  \int \ud^2 {\bf x} \, \rho({\bf x},t_i) (V({\bf x},a_f) - V({\bf x},a_i)) \nonumber \\
&=& \frac{1}{\beta} \, .
\end{eqnarray} 
%Therefore the behaviors of the mean works in the instantaneous jump limit is independent of processes.

\begin{figure}[t]
\begin{center}
%\vspace*{-3cm}
\includegraphics[scale=0.3]{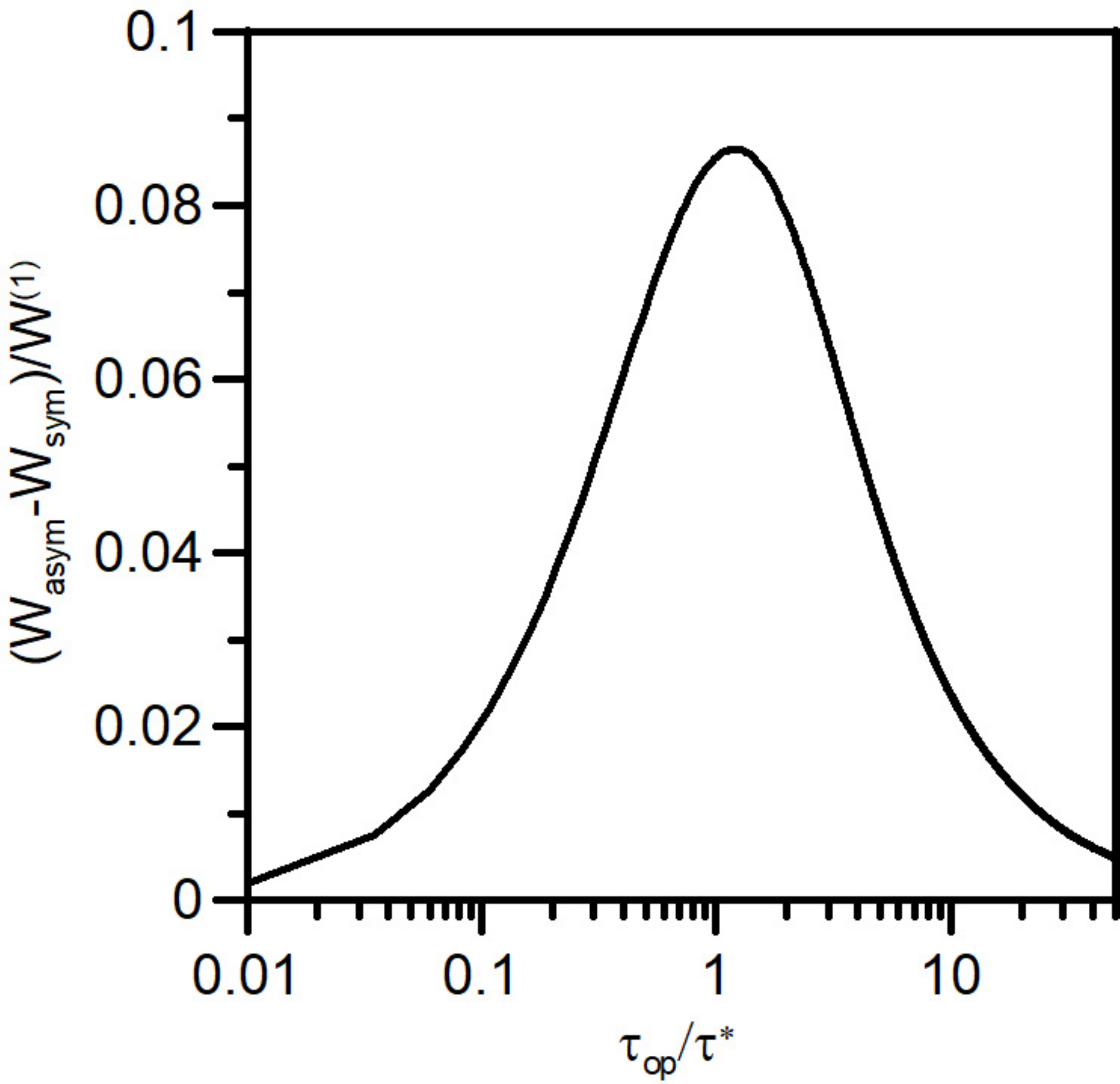}
\end{center}
\caption{
The difference between $W_{asym}$ and $W_{sym}$ is plotted as a function of the operation time 
$\tau_{op}=t_f - t_i$.
The peak is located around $\tau_{op} \sim \tau^*$
We define $\tau^*=\nu/a_0$.}
\label{fig:work_diff}
\end{figure}

The mean work in the asymmetric process (gray solid line) 
is always larger than that in the symmetric process (black solid line).
This indicates that irreversible contribution in the asymmetric process is always larger than that in the symmetric process. 
To understand this behavior, we should notice that Eq.\ (\ref{eqn:wirr_asym}) is reexpressed as 
\begin{eqnarray}
W^{(2)}_{asym} = 2 \int^{1/2}_{0} \ud \tau_1 \int^{\tau_1}_{0} \ud \tau_2 \, 
\dot{\bar{f}}_{\tau_2} 
\left[ 
 e^{-\frac{2\tau_{op} a_0}{\nu} \int^{\tau_1}_{\tau_2} \ud \tau_3 \, \bar{f}_{\tau_3} }
\frac{1}{2 \beta (\bar{f}_{\tau_2})^2} 
\right]
\dot{\bar{f}}_{\tau_1} \, .
\label{eqn:w2asymalta}
\end{eqnarray}
To obtain this expression, we have used 
that $\bar{f}_\tau$ and $\bar{g}_\tau$ defined by Eq.\ (\ref{eqn:opt_profg}) satisfy 
\begin{eqnarray}
\bar{g}_{\tau+1/2} = \bar{f}_\tau \, .
\end{eqnarray}
when $\tau_X = 1/2$.
Then one can see that $W^{(2)}_{asym}$ (\ref{eqn:w2asymalta}) is reproduced from $W^{(2)}_{sym}$ 
by replacing $\tau_{op}$ with $\tau_{op}/2$ in Eq.\ (\ref{eqn:wirr_sym_adim}). 
That is, the deformation of the harmonic potential in the asymmetric process is twice as fast as that in the symmetric process. 
Normally, the irreversible contribution is enhanced in rapid deformations and thus 
the mean work in the asymmetric process has more irreversible contributions than that in the symmetric process.

The difference of the mean works in the optimized symmetric and asymmetric processes disappears in the large $\tau_{op}$ 
and the instantaneous jump limits.
Therefore, we have to choose an appropriate operation time to maximize the difference.
As shown in Fig.\ \ref{fig:work_diff}, 
the difference is maximized when $\tau_{op}$ is chosen to be close to $\tau^*= \nu/a_0$, 
which characterizes the relaxation time of the Fokker-Planck equation.
The exact position of the peak is located at a bit larger than $\tau^*$.  
This small deviation from $\tau^*$ will be related to a hysteresis effect which is discussed in the next section.

\subsection{Change of Mean Energy}

\begin{figure}[t]
\begin{center}
%\vspace*{-3cm}
\includegraphics[scale=0.3]{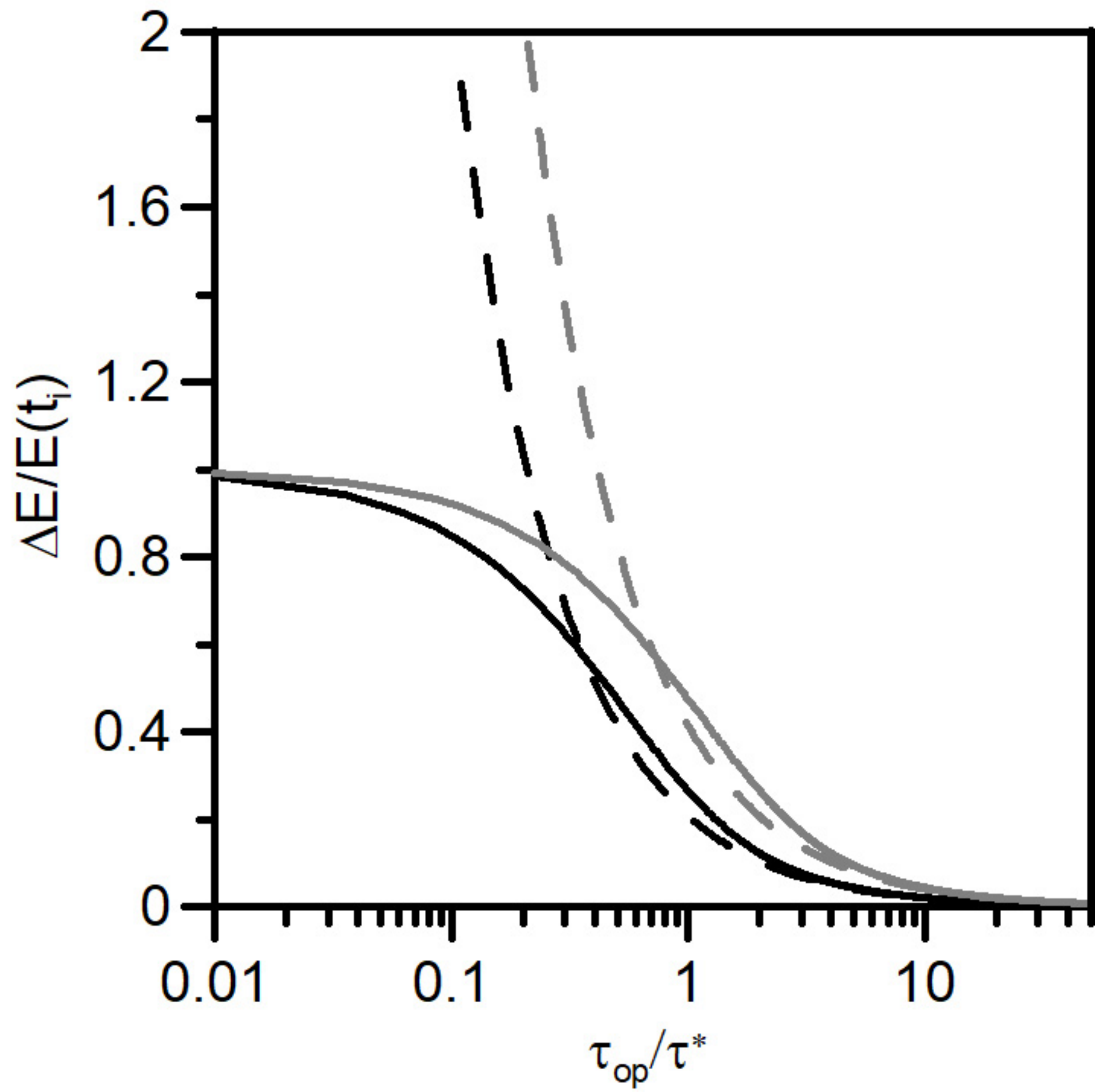}
\end{center}
\caption{The changes of the mean energy $\Delta E = E(t_f) - E(t_i)$ are plotted as functions of the operation time $\tau_{op}=t_f - t_i$.
The black solid and dotted lines are results in the symmetric process and represent the exact change of the mean energy  
and the change in the large $\tau_{op}$ limit, respectively.
The optimized control parameters are given by Eq.\ (\ref{eqn:app_protocol}).
The gray solid and dotted lines are results in the asymmetric process and represent the exact change of the mean energy 
and the change in the large $\tau_{op}$ limit, respectively.
The optimized parameters are given by Eq.\ (\ref{eqn:opt_profg}).
We define $\tau^*=\nu/a_0$.}
\label{fig:ener}
\end{figure}

\begin{figure}[t]
\begin{center}
%\vspace*{-3cm}
\includegraphics[scale=0.3]{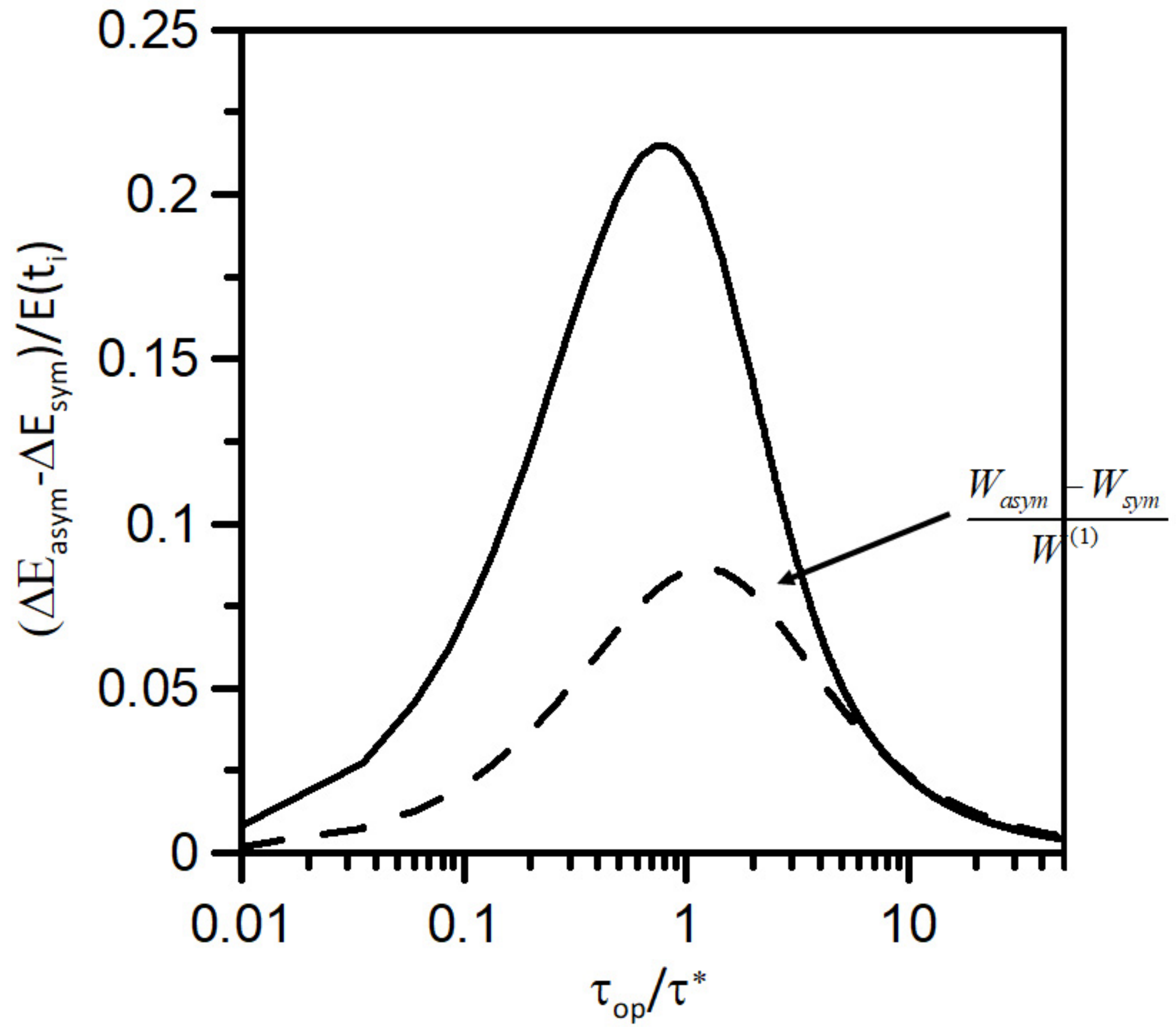}
\end{center}
\caption{
The difference between $\Delta E_{asym}$ and $\Delta E_{sym}$ is plotted as a function of the operation time 
$\tau_{op}=t_f - t_i$.
For the sake of comparison, the result of Fig.\ \ref{fig:work_diff} is shown by the dotted line. 
The peak of the solid line is smaller than $\tau^*$ but that of the dotted line is larger.  
We define $\tau^*=\nu/a_0$.}
\label{fig:ener_diff}
\end{figure}

The changes of the mean energy in the symmetric and asymmetric processes are shown as functions of the operation time $\tau_{op}$ in Fig.\ \ref{fig:ener}.
The black and gray solid lines represent the exact changes of the mean energy in the symmetric 
and asymmetric processes, respectively. 
The former is calculated from Eqs.\  (\ref{eqn:deltae_sym}) and the latter from Eqs.\ (\ref{eqn:deltae_asym}).
The black and gray dotted lines show the changes in the large $\tau_{op}$ limit which are 
given by Eqs.\ (\ref{eqn:e_largetau}) and (\ref{eqn:e_largetau_asym}), respectively.
We find that the asymptotic behaviors of the solid lines are approximately reproduced by the corresponding dotted lines. 
%These quantities vanish in the large $\tau_{op}$ limit which corresponds to the quasi-static isothermal process.

In this system, the mean energy in equilibrium is given by $1/\beta$ 
independently of the value of ${\bf a}_t$. 
Therefore the deviation of $\Delta E$ from zero can be used to characterize the deviation of the final state from equilibrium. 
Indeed, the change of the mean energy is reexpressed as 
\begin{eqnarray}
\Delta E = 
\int \ud^2 {\bf x} \, V({\bf x},{\bf a}_{t_f})
\left[ \rho({\bf x},t_f) 
- \frac{e^{-\beta V({\bf x},{\bf a}_{t_f}) } }{ {\cal Z} ({\bf a}_{t_f}) } \right] \, ,
\end{eqnarray}
where $V({\bf x},{\bf a}_{t_f})$ is the harmonic potential (\ref{eqn:harmonic_pot}).
The quantity in the bracket represents the deviation of $\rho({\bf x},t_f)$ from the corresponding thermal equilibrium distribution.
To obtain this, we have used 
\begin{eqnarray}
\int \ud^2 {\bf x} \, V({\bf x},{\bf a}_{t_i}) \frac{e^{-\beta V({\bf x},{\bf a}_{t_i}) } }{ {\cal Z} ({\bf a}_{t_i}) }  
=
\int \ud^2 {\bf x} \, V({\bf x},{\bf a}_{t_f}) \frac{e^{-\beta V({\bf x},{\bf a}_{t_f}) } }{ {\cal Z} ({\bf a}_{t_f}) } = \frac{1}{\beta}\, . 
\end{eqnarray}
From Fig.\ \ref{fig:ener}, one can see that $\Delta E$'s represented by the black and gray lines converge to zero in the large $\tau_{op}$ limit 
which corresponds to the quasi-static limit.

In the instantaneous jump limit $\tau_{op} \rightarrow 0$, as discussed below Eq.\ (\ref{eqn:w_instantaneous}), 
there is no enough time for the final state to evolve from the initial equilibrium state and thus 
\begin{eqnarray}
\lim_{\tau_{op}\rightarrow 0}  \Delta E^{sym} = \lim_{\tau_{op}\rightarrow 0}  \Delta E^{asym} = \frac{1}{\beta} \, .
\end{eqnarray}
This is consistent with the behaviors in Fig.\ \ref{fig:ener}.
We further observe that the gray lines are always larger than the black lines.
This is because, as was discussed in the mean works, 
the asymmetric process has larger irreversible contributions which give rise to the larger deviation of the final state from equilibrium.

The difference between $\Delta E_{asym}$ and $\Delta E_{sym}$ is shown in Fig.\ \ref{fig:ener_diff}.
For the sake of comparison, the result of Fig.\ \ref{fig:work_diff} is shown by the dotted line. 
We find that it is maximized when $\tau_{op}$ is chosen to be close to $\tau^*$.
Exactly speaking, however, the exact position of the peak is a bit smaller than $\tau^*$ 
and this behavior is 
different from that in the mean work which is a bit larger.
This difference will be related to a hysteresis effect: 
the change of the mean energy is obtained directly from the initial and final states 
while the mean work depends on the whole time evolution of the pseudo density matrix as seen from 
Eqs.\ (\ref{eqn:def_ener}) and (\ref{eqn:def_work}).
Because of this, 
the peak in the mean work appears in retard compared to that in the change of the mean energy.

\section{concluding remarks} \label{sec:concl}

In this paper, we developed the systematic expansion of the solution of the Fokker-Planck equation with degeneracy by generalizing the formulation developed in Ref.\ \cite{koide17}.
The Fokker-Planck equation describes the thermal relaxation of a Brownian particle confined in an external potential, 
which is deformed by changing control parameters.
The time derivative of these parameters are utilized as the expansion parameter of the perturbation theory. 
Differently from the theory in Ref.\ \cite{koide17}, the present theory is applicable to systems in arbitrary spatial dimensions and 
we obtained a new formula to calculate the mean work perturbatively which is applicable to the systems with degeneracy in the eigenvalues of the Fokker-Planck operator.
This formula enables us to study how the geometrical symmetry of the external potential affects thermodynamic description of Brownian motion.

The application of the derived formula depends on the degeneracy of systems.
To illustrate this, we considered the thermodynamic description of a Brownian particle confined in the two-dimensional harmonic potential. 
By changing the control parameters, the harmonic potential is monotonically compressed. 
Fixing the initial and final forms of the potential, we considered the symmetric and asymmetric processes.
The rotational symmetry of the harmonic potential in the two-dimensional plan is held in the former, but it is broken in the latter.
Perturbative calculations are affected by these deformation processes because 
the degeneracy of the expansion basis (foot and head states) depends on the symmetry of the potential. 

The optimized deformation processes of the harmonic potential are obtained by minimizing the mean works. 
The exact solutions have, however, not yet been known (see the discussion in Ref.\ \cite{koide17}) and thus 
we considered the optimization in the limit of the large operation time $\tau_{op}=t_f - t_i$. 
Using these approximated optimized processes,
we found that the mean works and the changes of the mean energy  in the asymmetric process are always larger than those in the symmetric process for any $\tau_{op}$, because 
the time scales of the deformation in the asymmetric process is shorter than that in the symmetric process.

When the operation time $\tau_{op}$ is very small, there is no enough time for the initial state to evolve.
Moreover the processes in the large $\tau_{op}$ limit converge to the quasi-static process. 
Therefore the symmetric and asymmetric processes show the same behaviors in these limits. 
To see the difference between the symmetric and asymmetric processes, we have to choose $\tau_{op}$ appropriately.
We calculated the difference of the mean works in the optimized symmetric and asymmetric processes as a function of $\tau_{op}$ 
and found that the difference is maximized when $\tau_{op} \sim \tau^* = \nu /a_0$ 
where $\nu$ is the coefficient of friction in the Fokker-Planck equation 
and $a_0$ denotes the initial value of the spring constant of the harmonic potential.

The similar difference can be calculated for the changes of the mean energy. 
We then found that the difference between the symmetric and asymmetric processes 
is maximized when $\tau_{op}$ is close to $\tau^*$ but is smaller than the corresponding value in the mean works.
This deviation may be related to a hysteresis effect: 
the change of the mean energy is obtained directly from the initial and final states 
while the mean work depends on the whole time evolution of the pseudo density matrix.
It is thus reasonable to consider that the maximum in the mean work appears in retard compared to that in the change of the mean energy.

In this theory, the solution of the Fokker-Planck equation is expanded with the foot and head states which are the eigenfunctions 
of the time-dependent Fokker-Planck operator.
These states form the bi-orthogonal system and thus the foot or head state itself is not necessarily normalized by one. 
As a consequence, there is an ambiguity to multiply a real factor to the foot and head states. 
See the factor ${\cal N}_{n,\greekbf{\alpha}_n} ({\bf a}_t)$ in Eq.\ (\ref{eqn:urho_u}).
Our perturbation theory should be invariant for this ambiguity. 
Indeed, the mean works and the changes of the mean energy in the harmonic potential do not depend on this additional real factor ${\cal N}_{n,\greekbf{\alpha}_n} ({\bf a}_t)$.
This invariance is however not yet shown for general external potential.

To apply the present formulation to non-linear potentials, 
we have to develop another perturbation theory to express the foot and head states 
in terms of those of an analytically solvable potential like the harmonic potential. 
This perturbation has not yet been studied.
The Floquet theory is known to be useful to solve periodic linear differential equations and 
provides a convenient method in quantum mechanics 
\cite{shirley}. 
This approach is applied to the Fokker-Planck equation \cite{caceres} but 
the corresponding optimization problems has not yet been investigated.
In the present paper, we assumed that the changes of the control parameters are given by smooth deterministic functions, but these changes 
can be stochastic in a microscopic time scale. For the optimization of stochastic control parameters, 
the minimization of the mean work will be replaced with the stochastic calculus of variation \cite{svm_review1,svm_review2}.
The applications to relativistic \cite{kk-rel,koide18,pal,parag} and quantum \cite{qse-koide} systems have not yet been investigated sufficiently.
These generalizations are left as future tasks.

\vspace*{1cm}
The author acknowledges the financial support by CNPq (303468/2018-1).
A part of the work was developed under the project INCT-FNA Proc.\ No.\ 464898/2014-5.

\appendix

\section{Properties of foot and head states} \label{app:1}

We should notice that the operator ${\cal H}_t$ can be expressed as 
\begin{eqnarray}
{\cal H}_t = \frac{1}{\nu \beta}  \sum_{i=1}^D {B}^\dagger_i {B}_i \, ,
\end{eqnarray}
where
\begin{eqnarray}
B_i &=& \partial_i + \frac{\beta}{2} (\partial_i V({\bf x},{\bf a}_t)) \, , \\ 
B^\dagger_i &=& - \partial_i + \frac{\beta}{2} (\partial_i V({\bf x},{\bf a}_t)) \, .
\end{eqnarray}
These operators satisfy the following commutation relations:
\begin{eqnarray}
[B_i, B^\dagger_j] &=& \beta (\partial_i \partial_j V({\bf x},{\bf a}_t)) \, , \label{eqn:bbcom} \\
{\protect [B_i, B_i]} &=& 0 \, , \\
{\protect [B^\dagger_i, B^\dagger_j]} &=& 0\, .
\end{eqnarray}
Then the eigenvalues are shown to be non-negative,
\begin{eqnarray}
\lambda_n (t)= \int \ud^D {\bf x} \, u^*_{n,\greekbf{\alpha}_n} ({\bf x},t) {\cal H}_t u_{n,\greekbf{\alpha}_n} ({\bf x},t) = \frac{1}{\nu \beta} \sum_{i=1}^D |B_i  u_{n,\greekbf{\alpha}_n} ({\bf x},t) |^2 \ge 0 \, .
\end{eqnarray}
Thus, without loss of generality, the eigenvalues are ordered as 
\begin{eqnarray}
0 = \lambda_0 (t) < \lambda_1 (t) < \lambda_ 2 (t) \cdots \, .
\end{eqnarray}

The foot state with $\lambda_0 = 0$ is given by solving $B_i u_{0,\greekbf{\alpha}_0} = 0$ 
and we find 
\begin{eqnarray}
\underline{\rho}_{0,{\bf 0}} ({\bf x},{\bf a}_t) = \sqrt{{\cal N}_{0,{\bf 0}}({\bf a}_t)}e\frac{1}{\sqrt{{\cal Z} ({\bf a}_t)}} e^{-\beta V({\bf x},{\bf a}_t)} \, ,
\end{eqnarray}
where
\begin{eqnarray}
{\cal Z} ({\bf a}_t) = \int \ud^D {\bf x} \, e^{-\beta V({\bf x},{\bf a}_t)} 
\, .
\end{eqnarray}
One can see that this is the stationary solution of the Fokker-Planck equation when $V$ is not time-dependent explicitly.
There is no degeneracy in the stationary solution and thus we set $\greekbf{\alpha}_{0} = {\bf 0}$.
The corresponding head state is easily found by using Eq.\ (\ref{eqn:ou1}), 
\begin{eqnarray}
\overline{\rho}_{0,{\bf 0}} ({\bf x},{\bf a}_t) = \frac{1}{\sqrt{{\cal N}_{0,{\bf 0}}({\bf a}_t)}}\frac{1}{\sqrt{{\cal Z}({\bf a}_t)}} \, .
\end{eqnarray}

It should be noted that $B_i$ and $B^\dagger_i$ are not the lowering and raising operators in general because 
the right-hand side of Eq.\ (\ref{eqn:bbcom}) is not necessarily constant.

\end{document}